\documentclass[sigconf]{acmart}
\usepackage{bbm}
\usepackage{comment}
\usepackage{multirow}

\AtBeginDocument{%
  \providecommand\BibTeX{{%
    \normalfont B\kern-0.5em{\scshape i\kern-0.25em b}\kern-0.8em\TeX}}}

\usepackage{xspace}
\usepackage{amsmath}
\usepackage{bbold}
\usepackage{dsfont}

\newcommand{\cb}{}

\definecolor{darkgreen}{rgb}{0.078,0.667,0.016}

\usepackage{subcaption}

\newcommand{\sys}{QarSUMO\xspace}

\newcommand{\para}[1]{\noindent \textbf{#1 }}
\newcommand{\outline}[1]{}

\newcommand{\eg}{\textit{e.g.}\xspace}

\newcommand{\san}[1]{\textcolor{purple}{Sanjay:#1}}
\newcommand{\ste}[1]{\textcolor{blue}{Stefano:#1}}
\newcommand{\yk}[1]{\textcolor{orange}{Ke:#1}}

\copyrightyear{2020}
\acmYear{2020}
\setcopyright{acmlicensed}\acmConference[SIGSPATIAL '20]{28th International Conference on Advances in Geographic Information Systems}{November 3--6, 2020}{Seattle, WA, USA}
\acmBooktitle{28th International Conference on Advances in Geographic Information Systems (SIGSPATIAL '20), November 3--6, 2020, Seattle, WA, USA}
\acmPrice{15.00}
\acmDOI{10.1145/3397536.3422274}
\acmISBN{978-1-4503-8019-5/20/11}

\begin{document}
\sloppy

\title{QarSUMO: A Parallel, Congestion-optimized Traffic Simulator}


\author{Hao Chen}
\authornote{Both authors contributed equally to this research.}
\email{chenhao@hbku.edu.qa}
\affiliation{%
  \institution{University of Science and Technology of China \& Qatar Computing Research Institute, HBKU}
}

\author{Ke Yang}
\authornotemark[1]
\email{yke@hbku.edu.qa}
\affiliation{%
  \institution{Tsinghua University \& Qatar Computing Research Institute, HBKU}
}

\author{Stefano Giovanni Rizzo}
\email{strizzo@hbku.edu.qa}
\affiliation{%
  \institution{Qatar Computing Research Institute, HBKU}
}

\author{Giovanna Vantini}
\email{gvantini@hbku.edu.qa}
\affiliation{%
 \institution{Qatar Computing Research Institute, HBKU}
}

\author{Phillip Taylor}
\email{phillipt@princeton.edu}
\affiliation{%
 \institution{Princeton University}
}

\author{Xiaosong Ma}
\email{xma@hbku.edu.qa}
\affiliation{%
  \institution{Qatar Computing Research Institute, HBKU}
}

\author{Sanjay Chawla}
\email{schawla@hbku.edu.qa}
\affiliation{%
  \institution{Qatar Computing Research Institute, HBKU}
}

\renewcommand{\shortauthors}{Hao and Ke, et al.}

\begin{abstract}
\label{sect:abstract}
Traffic simulators are important tools for tasks such as urban planning and transportation management. 
Microscopic simulators allow per-vehicle movement simulation, but require longer simulation time. 
The simulation overhead is exacerbated when there is traffic congestion and most vehicles move slowly. 
This in particular hurts the productivity of emerging urban computing studies based on reinforcement learning, where traffic simulations are heavily and repeatedly  used for designing policies to optimize  traffic related tasks.

In this paper, we develop QarSUMO, a parallel, congestion-optimized version of the popular SUMO open-source traffic simulator. 
QarSUMO performs \textit{high-level parallelization} on top of SUMO, to utilize powerful multi-core servers and enables future extension to multi-node parallel simulation if necessary. 
The proposed design, while partly sacrificing speedup, 
makes QarSUMO compatible with future SUMO improvements. 
We further contribute such an improvement by modifying
the SUMO simulation engine for congestion scenarios where
the update computation of \textit{consecutive and slow-moving
vehicles} can be simplified.
We evaluate QarSUMO with both real-world and synthetic road network and traffic data, and examine its execution time as well as simulation accuracy relative to the original, sequential SUMO. 
\end{abstract}





\maketitle
\section{Introduction}
\label{sec:introduction}
The design of road infrastructures and the planning of traffic control are 
challenging tasks,  often requiring 
complex  modeling to verify and analyze feasible solutions safely and efficiently. 
Road traffic simulators are specialized software widely used in research  and practice
and make it possible to recreate real-world scenarios, such as the congestion surrounding a stadium after a sport event, or the traffic flow in a major city during rush hours. 
Through simulators it is possible to experiment with different ``what-if'' scenarios, predicting their outcome with a certain degree of accuracy, as well as validating different solutions to mitigate unwanted effects.

There are three main types of traffic simulators: \textit{macroscopic}, \textit{mesoscopic}, and \textit{microscopic}~\cite{sokolowski2011principles}. Macroscopic simulators focus on generating higher-level aggregate traffic statistics,  
while mesoscopic ones work at the level of groups of vehicles. 
The focus of our work is on microscopic simulators, where individual vehicles in a network are modeled.
The demand for such detailed simulation is
increasing, as they enable the evaluation of fine-grained traffic control policies, such as the impact of a new traffic signal system on rush hour traffic with each individual vehicle adopting dynamic routing.
However, the main challenge with microscopic simulation is that such detailed computation is time-consuming - as the scale of the simulation grows, the per-timestep simulation overhead increases accordingly, possibly preventing quick and timely responses as requested under real time operating conditions.
The modeling of traffic congestion further dramatically increases the simulation time.
As common microscopic simulation performs discreet object simulation by moving each vehicle at preset physical (wall) timestep intervals, such as 0.5 second, in congested scenarios  the number of vehicles and interactions between them grows, resulting in poor performance. 
Unfortunately, traffic congestion scenarios are important study subjects and natural optimization targets, making their costly simulation necessary for users. 


The above challenges are highlighted in new use scenarios of traffic simulation. 
In particular, traffic simulators and their performance have played a central role in the recent trend of applying Reinforcement Learning (RL) techniques in urban computing~\cite{wei2019survey,kheterpal2018flow,2017_PG_and_QLearning_intersection_image,rizzo2019time,2016_multiagent_reinforcement_learning_coordination_graphs}.
All these approaches rely heavily on the repeated simulation of thousands of \textit{episodes}, in which an agent, such as a traffic light planner, explore a huge number of combinations of actions in order to  learn an optimal policy.
It follows that in order to enable such techniques for traffic control, the simulations have to run at a speed that is orders of magnitude faster then the real time.  
This is one of the reasons why RL has been so far applied only on single intersections~\cite{intellilight,rizzo2019time} or on simple grid-like networks~\cite{wei2019presslight,2016_multiagent_reinforcement_learning_coordination_graphs}. 
It is clear that the future success of these approaches is strongly tied to the efficiency and scalability of the underlying traffic simulators.

In this paper, we tackle the scalable and efficient traffic simulation challenge by enabling parallel simulation. 
Rather than building new systems, we choose to improve the popular open-source SUMO simulation software~\cite{lopez2018microscopic}, which has already established a sizable user community (with an annual user conference~\cite{sumoconfwebsite}).
In exchange for its rich features (such as detailed microscopic simulation, support of different traffic and routing modes, and powerful interfaces), SUMO suffers longer execution times compared to lightweight alternatives~\cite{tang2019cityflow, ramamohanarao2016smarts, fu2019building}. 
For example simulating one hour of traffic involving a million vehicles traversing a city, it can take SUMO to processing time non responding to requested needs~\cite{krajzewicz2002sumo}.

Our proposed system, \sys, adds high-level parallelism on top of SUMO simulation, by adding network partitioning and inter-partition vehicle state synchronization along the border edges. This approach retains SUMO internal designs, allowing orthogonal optimizations to be easily incorporated.
Considering that powerful multi-core servers today provide considerable hardware parallelism, and that typical road network data could easily fit into a single node's main memory, we focus in this work on single-server parallel execution using multiple threads. 
\sys, however, adopts MPI communication~\cite{gabriel2004open} and can be easily extended to multi-node parallel execution in the future. 

We then focus on SUMO internal optimizations to speed up traffic simulation under congestion. 
The intuition here is that vehicles move very slowly, maintaining their close distances, while events like lane change become more rare.  Therefore, when we temporarily reduce the simulation granularity and model a consecutive sequence of vehicles within the same lane,
it does not significantly affect the simulation outcome.
Based on dynamically monitored congestion level, length of road segment, and distance to the next road junction, \sys judiciously applies such \textit{group simulation} to save simulation time while retaining reasonable accuracy relative to the SUMO baseline. 

We evaluate \sys with both synthetic grid networks and a real road network of the Doha Corniche area.
This area hosts arterial traffic linking key government agencies, major tourist attractions, and many hotels, making it a strategic transportation management target in preparation for the Qatar 2022 FIFA World Cup. 
Our results show that \sys is able to bring an order of magnitude performance improvement for SUMO simulations, while maintaining reasonably high accuracy. 
\sys will be released as open-source software. 

The rest of the paper is organized as follows. 
Section~\ref{sec:background} gives background information on SUMO. 
Section~\ref{sec:meta-parallelization} and Section~\ref{sec:congestion-optimization} describes \sys design on parallel simulation and congestion simulation optimizations, respectively. 
Section~\ref{sec:evaluation} reports evaluation and Section~\ref{sec:related-work} discusses related work. 
Finally, Section~\ref{sec:conclusion} summarizes the paper as well as potential future work.

\section{Background: SUMO}
\label{sec:background}
\subsection{Microscopic Traffic Simulation}
\label{subsec:micro}
Most microscopic simulators model
the complete traffic ecosystem including the road network with lanes, traffic signals, routing strategy, and the kinematics of vehicles. 
The movement of vehicles are modeled using a {\it car following model (CFM)} and a {\it lane changing model (LCM)}~\cite{sokolowski2011principles,gazis}.
At the most basic level, a CFM models the acceleration  $a_{i,t+1}$  of a vehicle $i$ at time step $t+1$ as:  
\[
a_{i,t+1} = f(v_{i,t},v_{d},v_{i-1,t})
\]
Here $v_{i,t}$ is the speed of the vehicle $i$ at time $t$, $v_{d}$ is the desired speed, and $v_{i-i,t}$ is the speed of the immediate vehicle ahead. 
When the vehicle in question sees no immediate traffic ahead or faces a traffic signal, the vehicle ahead is replaced by a fictitious vehicle or an impeding object, respectively. 
The function $f()$ depends upon the specific CFM.

The LCM, on the other hand, is a binary decision-making function that decides whether it is appropriate for a vehicle to change lanes at a given time step:
\[
\mathds{1}_{ > 0}(i,t+1) =
g(a_{i,t},a_{i-1,t},a_{i+1,t})
\]
Note that unlike the CFM, LCM depends not only on the vehicle in front ($v_{i-1,t}$), but also the vehicle that is behind in the target lane ($v_{i+1,t}$). 
Here the function $g()$ depends upon the specific LCM that is being used.

From a distributed computing perspective, the key challenge is to synchronize the computation
of both $f()$ and $g()$ across partitions. 
For example, with CFM, if the current vehicle and
the vehicle immediately ahead are in different partitions then provisions have to be made
to communicate the information across partitions. 
This is the key reason that parallelizing a microscopic simulator in a manner that is accurate and efficient at the same
time is non-trivial.

There are several variations of both CFM and LCM in the literature~\cite{sokolowski2011principles,gazis}. 
One advantage of using open-source simulators like SUMO (to be described below) is that these alternative models can be implemented and tested by users.  


\subsection{SUMO Overview}
\label{subsec:sumo}
SUMO \cite{lopez2018microscopic} is an open source, freely available microscopic simulation software.
It offers a great number of features and extensions, and is currently used worldwide for traffic research~\cite{sumo:projects}.

SUMO can read road networks from multiple popular formats, including OpenStreetMap, OpenDRIVE, Shapefile, Vissim, and MATSim.
Traffic demand can be modeled in SUMO as specified routes, individual trips from origin to destination, or high-level flows generating periodic trips. 
Routes can be generated statically or dynamically and vehicles are distributed among the routes with different strategies, for example, according to predefined statistics on turning ratios or detectors data. 
SUMO can also simulate multi- and inter-modal trips: each trip can use single- or multi- transportation mode. For example each individual can  move by walking, riding a vehicle, or using public transport, and also transferring among different modes, \eg, by car and then by rail.

Among the many advanced features of the vehicle interactions, SUMO supports different strategies for lane-changing, driving impatience parameters for overtaking, multiple pollutants emission models, and the configuration of onboard bluetooth and wifi devices, to simulate wireless sensor detection.
Finally, a very fine-grain level of details can be managed and analyzed on each single vehicle and road segment. 
For electric vehicles, for example, parameters such as battery capacity, drive efficiency, and the minimum velocity to start charging can be specified, along with charging station locations in a road network.

SUMO allows external programmatic interactions through two different interfaces: TraCI (Traffic Control Interface) and Libsumo. Both these interfaces provide the capability to retrieve values of any simulated objects, modify most of the objects settings, and have control on the running simulation while it is running. 
The TraCI interface\footnote{\texttt{https://sumo.dlr.de/docs/TraCI.html}} is a TCP-based client/server API and the most common way of interacting with SUMO. The main advantage of TraCI is that it abstracts from any specific language or platform, leaving users with full flexibility. 
On the flip side, TraCI brings high communication overhead, an issue particularly in partitioned simulations~\cite{arroyo2018new}.  
Libsumo on the other hand provides a much more efficient coupling, exposing the same interface methods of TraCI as C++ static functions, though with few limitations\footnote{\texttt{https://sumo.dlr.de/docs/Libsumo.html}} and less mature support than with TraCI.

\begin{figure}[h!]
  \centering
  \includegraphics[width=0.45\textwidth]{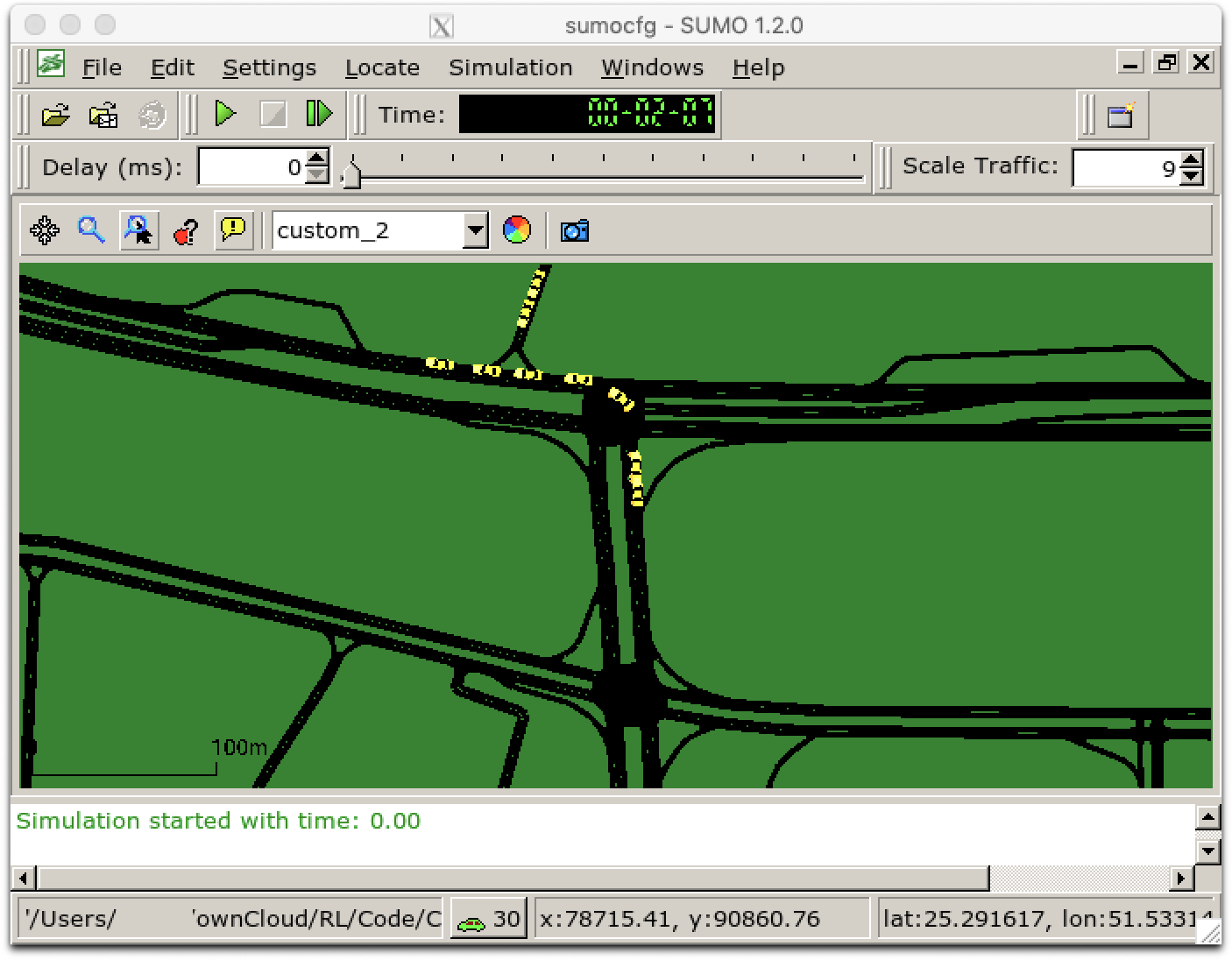}
  \caption{Screenshot of a running simulation in SUMO.}
\label{fig:screenshot}
\end{figure}

Finally, SUMO provides user-friendly GUI for the user to interact with the simulation and visualize intermediate results, thus facilitating the understanding and analysis of traffic scenarios.
Figure~\ref{fig:screenshot} gives a sample SUMO screenshot that shows a running simulation on a real road network. The map network can be navigated and zoomed, while specific objects can be selected to monitor related parameters, such as the current route of a vehicle or its velocity.

\subsection{Key Entities and Actions in SUMO}
\label{subsec:sumo-ds}
We briefly introduce the key objects and important data structures in SUMO, before we move to \sys design. 

In order to simulate traffic, SUMO keeps track of real traffic networks as well as moving entities using its internal representations. 
Major objects concerned in this paper's discussion include:
\begin{itemize}
\vspace{-3pt}
    \item \textit{Junction}: a junction (node) is a single point where at least one road segment starts or ends. Junctions may have traffic lights, individually controlled by configurable signal transition policies. 
    \item \textit{Edge}: an edge represents a one-way, uninterrupted road segment between two road junctions, each with its speed limit and potentially containing multiple \textit{lanes}. Together, edges and junctions define a traffic network. Note that at intersections, we also have arch-shaped or straight lanes connecting the corresponding lanes of intersecting edges.
    \item \textit{Vehicle}: vehicles are the main simulation objects and contain many attributes, such as the current speed, position, and acceleration. They are stored in a per-lane array, for their successive updates.  
\end{itemize}

The simulation timestep is configurable, with SUMO default set at 0.5 second. 
At each timestep, SUMO examines the vehicles in each lane sequentially, consistent with the natural vehicle following behavior in real world. 
For each vehicle, SUMO adjusts its speed using the selected CFM, considering factors such as the distance with the vehicle ahead, speed limit, distance to junction, acceleration, etc.
A vehicle may initiate other events, such as lane change, as required by routing or enabled as probabilistic actions. 

To simulate real-world driving scenarios, SUMO also inserts configurable randomness into the vehicles' behavior. 
As such non-deterministic executions make it more difficult for QarSUMO to compare with the original SUMO for simulation accuracy, we disable such randomness insertion in our evaluation. 

\section{Meta-parallelization of SUMO}
\label{sec:meta-parallelization}

\subsection{\sys Architecture Overview}

\begin{figure}[htb]
 \centering
 \includegraphics[width=0.45\textwidth]{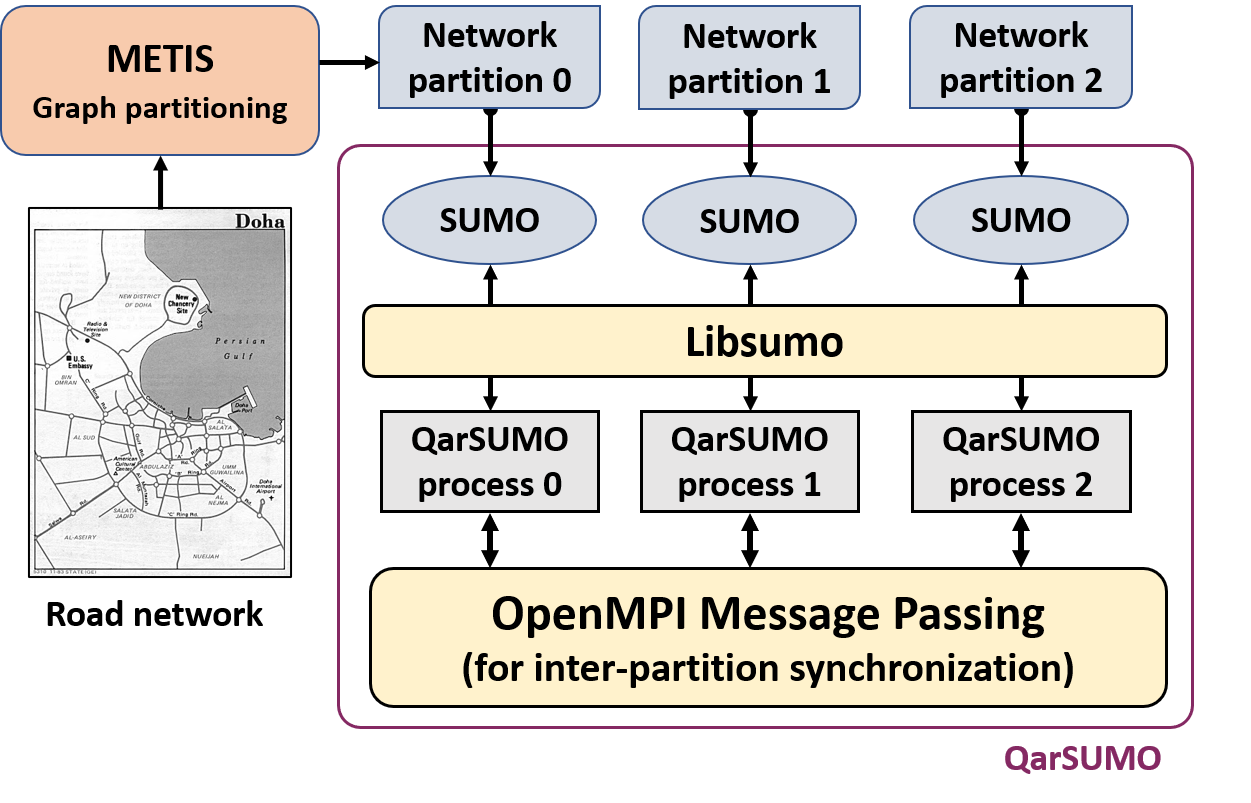}
 \vspace{-3mm}
 \caption{QarSUMO Architecture}
 \label{fig:qarsumo_architecture}
 \vspace{-3mm}
\end{figure}
\sys enables parallel simulation of SUMO by partitioning the road network (and the corresponding traffic). 
\sys adopts a fully distributed design, with no central coordinator.
Each partition is processed by a \sys process, who performs local updates, and exchanges vehicle states on \textit{cut edges} (the edges spanning two neighboring partitions) at the end of each simulation timestep. 

Figure~\ref{fig:qarsumo_architecture} illustrates the \sys architecture, using a simple example of 3-partition execution. 
\sys adopts the widely used METIS graph partitioning tool~\cite{karypis1998fast} to generate desired road network partitions. 
Each \sys process runs one instance of sequential SUMO simulation on its own network partition. 
This leaves all the communication and synchronization to the \sys processes, using OpenMPI~\cite{gabriel2004open} for message passing. 
The SUMO instance invoked within each \sys process, meanwhile, retains their sequential simulation workflow. 

\sys processes only need to interact with their corresponding SUMO instance at timestep granularity and the interaction is limited to only the traffic on border edges (along with their associated junctions). 
This is done via the standard Libsumo C++ interfaces. 
Details are to be given in Section~\ref{subsec:sync}. 

This design allows \sys's parallel execution to stay high-level and relatively independent of the SUMO sequential simulation implementation. 
As SUMO itself is been increasingly adopted and professionally maintained, such decoupled design allows \sys to easily accommodate new SUMO optimizations or upgrades. 
Our congestion optimization technique (to be presented in Section~\ref{sec:congestion-optimization}, for example, is one such enhancement that does modify SUMO internal implementation). 

Also, \sys's design is made considering potential future SUMO's own parallelization. 
Likely due to growing user demands, recently the SUMO team is adding its own multi-threaded parallel implementation. 
However, the work does not seem to be complete yet, and our evaluation finds the current SUMO parallel execution often producing longer simulation time than the sequential version, due to heavy inter-thread synchronization.
However, with potential future versions achieving reasonable multi-threaded execution efficiency, \sys's MPI-based meta-parallelization could provide additional speedup on top of it.
For example, if SUMO is able to scale out to 4 or 8 threads doing parallel simulation using shared memory, on a 32-core server \sys could run 8 or 4 such SUMO instances, each working on a network partition and invoked by a \sys process still communicating using message passing.  

Finally, MPI-based message passing design easily extends to multi-node parallel execution. 
In this paper we focus on single-node evaluation, considering our moderate sub-city scale road networks and users' preference of avoiding cluster setup.
Meanwhile, \sys has no problem utilizing multiple server nodes, when needed in processing larger networks or traffic volumes.

\subsection{Network Partitioning}
\label{subsec:partitioning}
For parallel simulation, \sys needs to first partition the road network.
It adopts \textit{vertex partitioning}, so that each junction is uniquely assigned to one partition.
If both ends of an edge belong to the same partition, then the edge becomes an \textit{internal edge} there.
Otherwise, the edge sits across two partitions and becomes a \textit{border edge} (cut edge).

Graph partitioning is a relatively mature field and \sys chooses to leverage the popular METIS tool~\cite{karypis1998fast}, which supports balanced partitioning with a variety of optimization objectives. 
\sys first extracts the road graph from the input traffic network, then partitions the graph using METIS, and finally converts the original network according to the partitioning results into corresponding network partitions. 
Compared to the simulation time (especially with iterative episode simulation for reinforcement learning training), the time spent on such preprocessing is minor.
For example, it takes METIS 0.01 second to partition the networks used in our evaluation (details in Section~\ref{subsec:setup}). 

Therefore, for long-running experiments simulating consistently imbalanced traffic (with some of the road segments much busier than others), it is worthwhile to factor in the traffic distribution to periodically re-generate network partitions in a load-balanced manner. To this end, \sys supports optional enhanced network partitioning by taking into account the routing information of vehicles (typically stored in the 
route file for SUMO simulation).

More specifically, METIS balances graph partitions by total vertex weight. 
Therefore \sys first traverses the static routing paths of all vehicles to be simulated and count the per-edge accesses. 
Then for each junction $v$, we calculate its traffic-aware vertex weight $w_v$ as:
\begin{align}
w_v' = \sum\limits_{i=1}^n {C_{e_i}L_{e_i}}, \\
w_v = \frac{1}{|V|} \sum\limits_{i=1}^{|V|} {w_{v_i}'} + w_v'
\end{align}

Here $e_i$ enumerates $v$'s incident edges, while
$C_{e_i}$ and $L_{e_i}$ are the access count and length of edge $e_i$, respectively.  
\cb{All junctions are assigned $\frac{1}{|V|} \sum\limits_{i=1}^{|V|} {w_{v_i}'}$ (the average weight over all vertices) as a base weight, so that they can be properly involved in partitioning even with no expected traffic.} 

With METIS partitioning assigns vertices (junctions in our case) to partitions. 
For parallel processing, the junctions connected by a border edge also need to be replicated at both partitions. 
For such a junction, we call its copy at the partition it is assigned to by METIS the \textit{primary junction} and the replicated one its \textit{shadow junctions}.

In addition to generating partitions with balanced total edge weight, METIS allows an additional user-specified optimization goal. 
\sys elects to minimize edge cut, as the communication overhead is heavily influenced by the number of border edges (more details on communication in Section~\ref{subsec:sync} next). 

\begin{table}[htb]
\begin{tabular}{|c|c|c|c|c|c|}
\hline
 Network/Partition   & 2 & 4  & 8  & 16 & 32 \\ \hline
Corniche   & 0.43\% & 1.01\% & 2.67\% & 4.54\% & 8.00\% \\
Cologne  &  0.07\% &  0.17\% & 0.33\%  &  0.77\% &  1.04\% \\
Grid   &  0.39\% &  1.13\% & 2.68\%  &  6.51\% &  11.02\% \\ \hline
\end{tabular}
\caption{Ratio of border edges}
\label{table:border_edge_number}
\vspace{-5pt}
\end{table}
Table~\ref{table:border_edge_number} demonstrates sampling partitioning results of METIS.
Less than 1\% of edges are border edges when we have only two partitions.
As expected, as we increase the number of partitions, both networks have growing percentage of edges becoming border edges.

\subsection{Inter-Partition Synchronization}
\label{subsec:sync}

\begin{figure}[!h]
 \centering
 \includegraphics[width=0.45\textwidth]{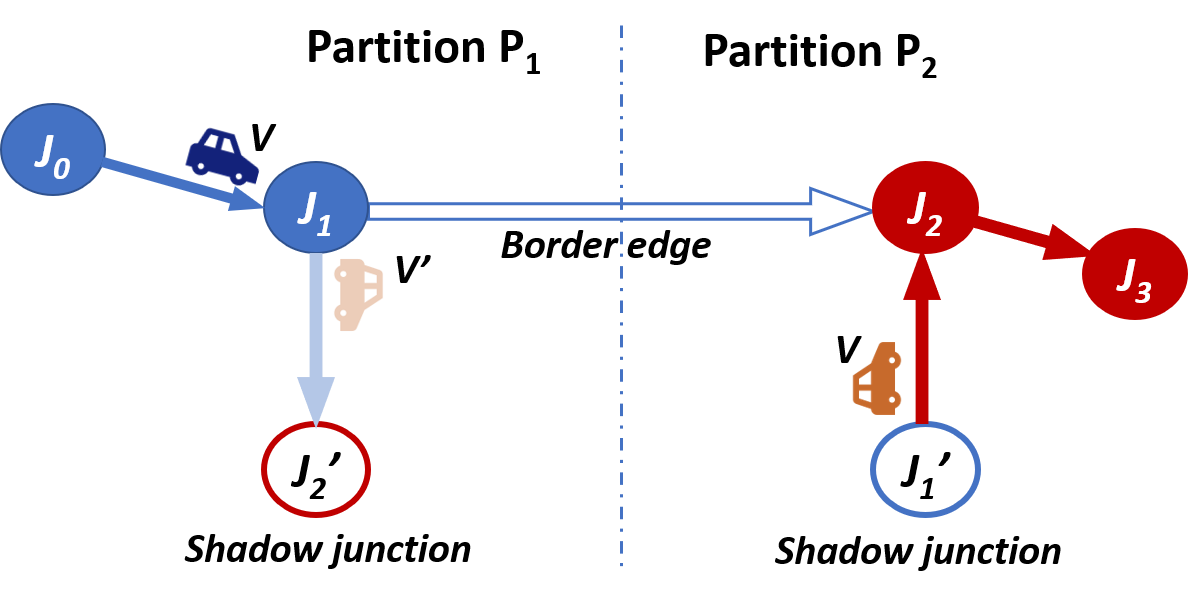}
 \caption{Synchronization between two partitions in QarSUMO}
 \label{fig:sync}
 \vspace{-5mm}
\end{figure}

At the end of each simulation timestep, \sys performs vehicle/lane state synchronization along the border edges. 
This is done by the replication of both road network and vehicle data traversing the border edges. 

As mentioned earlier, junctions are replicated across partitions. 
A border edge, accordingly, is also replicated. 
Figure ~\ref{fig:sync} illustrates a directed border edge, $J_1J_2$, sits across two partitions $P_1$ and $P_2$, where $J_1$ and $J_2$ are assigned, respectively. 
The shadow junctions, $J_1'$ and $J_2'$, reside at the opposite side.

In \sys's partitioned network, the border edge $J_1J_2$ does not exist. Instead, we have the replicated border edge $J_1J_2'$ in $P_1$, and $J_1'J_2$ in $P_2$. 
Considering that the traffic \textit{ahead} of a certain vehicle has far more influence on its movement than the traffic behind it, we consider the destination partition has more information to dictate a vehicle's state computation. 
Therefore, the edge $J_1'J_2$ (in $P_2$) is the \textit{primary edge} here, as $P_2$ hosts all the incident edges of $J_2$. 
while $J_1J_2'$ becomes its \textit{shadow edge}. 
The border edge of opposite direction, $J_2J_1$, has symmetric arrangements. 

When a vehicle $V$ moves from junction $J_0$ to $J1$, continuing via $J_1J_2$ (conceptually) toward $J2$, it emerges first in $P1$ on the shadow edge $J_1J_2'$. 
At the end of that timestep, the bi-directional communication between $P_1$ and $P_2$ will lead to the insertion of $V$'s replica in $P_2$, on edge $J_1'J_2$. 
Since this edge is a primary edge, the vehicle replica on it now becomes the \textit{primary vehicle}, with the original copy in $P_1$ reduced to be its \textit{shadow vehicle}.
From that point, $P_2$ takes over the main control of $V$'s simulation, and at the end of each timestep, while $V$ is still on $J_1'J_2$, sends updates to $P_1$. 

Note that the \sys process in charge of $P_1$ cannot simply drop $V$ from the shadow edge, which would affect the update of newly inserted vehicles after it.
Instead, it performs its local updates, while adjusting it after getting updates from $P_2$, allowing traffic situation to propagate from $J_2$ to $J_1$ along this border edge. 
At the end, when $V$ arrives in $J_2$ and moves on to an internal junction $J_3$, $V$'s shadow vehicle is removed from the shadow edge and disappears from $P_1$. 
Care is taken to deal with corner cases such as $V$ immediately returns to $P_1$ via another border edge. 

For cross-partition synchronization, at the end of each simulation timestep, a \sys process retrieves vehicle state updates concerning border edges from the underlying SUMO instance, using the aforementioned Libsumo interfaces. 
It aggregates such per-vehicle data to be sent to the same destination partition into a single MPI message.
All \sys processes use the \texttt{MPI\_Alltoall} collective communication call to simultaneously scatter/gather updates efficiently.
Finally, they each applies the appropriate update for vehicle insertion and shadow vehicle state update, again via Libsumo.

\section{Computation Optimization under Congestion}
\label{sec:congestion-optimization}

\subsection{Optimization Rationale}
\label{subsec:rationale}
As mentioned earlier, traffic simulation is especially time-consuming when simulating vehicles under traffic congestion. 
As vehicles move much slower compared with under light traffic, it takes much more simulation timesteps for them to move the same distance. 
Close examination of SUMO implementation reveals that the computation follows the same workflow regardless of congestion situations: each vehicle calculates its speed ahead according to the selected car following model, only to be severely constrained by the vehicle immediately ahead. 
It nevertheless follows its entire procedure, including spending considerable computation on checking lane change opportunities. 

In real life, cars stuck in traffic have little choice but to follow the car ahead in lock step, and drivers have little motivation to change lanes for the sake of moving faster. 
Therefore, under heavily congested traffic, we are spending a lot of cycles on simulating scenarios with not much happening. 
The intuition here is that one could probably achieve very similar simulation outcome when reducing the simulation granularity. 

We first explored enlarging the temporal granularity by increasing the simulation timestep, \eg, using larger timestep values than the SUMO default of 1 second. 
However, we found that this approach significantly impacts simulation accuracy when applied globally. 
To have varied simulation timesteps at different regions according to the traffic congestion level, on the other hand, significantly increases software complexity. 
In particular, SUMO has other timestep settings, such as "action timestep", which is used to recompute vehicle's speed and lane-changing decisions.
The interaction of these related parameters making it difficult to adjust the temporal granularity of simulation.  
Things become even more challenging when we consider \sys's parallelization, where the synchronization (described in the previous section) is built upon having consistent simulation timestep across all partitions and border edges.  

We then investigated varying the spatial granularity. 
Here we have two choices. 
One is to replace a sequence of vehicles packed under heavy congestion into a long ``super-vehicle'', whose movement is replicated to its member vehicles till its decomposition (due to dissolving congestion or approaching intersection). 
The other is to create a virtual group of such a vehicle sequence, make the first vehicle in the sequence a \textit{leader}, while skipping most of the simulation computation steps for the \textit{followers} by directly \textit{copying} leader simulation updates. 

Our examination of the SUMO simulation processing finds the second approach more lightweight and easier to implement. 
It also reserves each vehicle's individual examination, making it easier to address their individual situations. 
Less modification to SUMO data structures are needed (mostly extensions to the lane and vehicle object structures to add additional attributes). We discuss the design of this virtual grouping (``grouping'' for short) approach in the rest of this section. 

\begin{figure}[htb]
 \centering
 \includegraphics[width=0.45\textwidth]{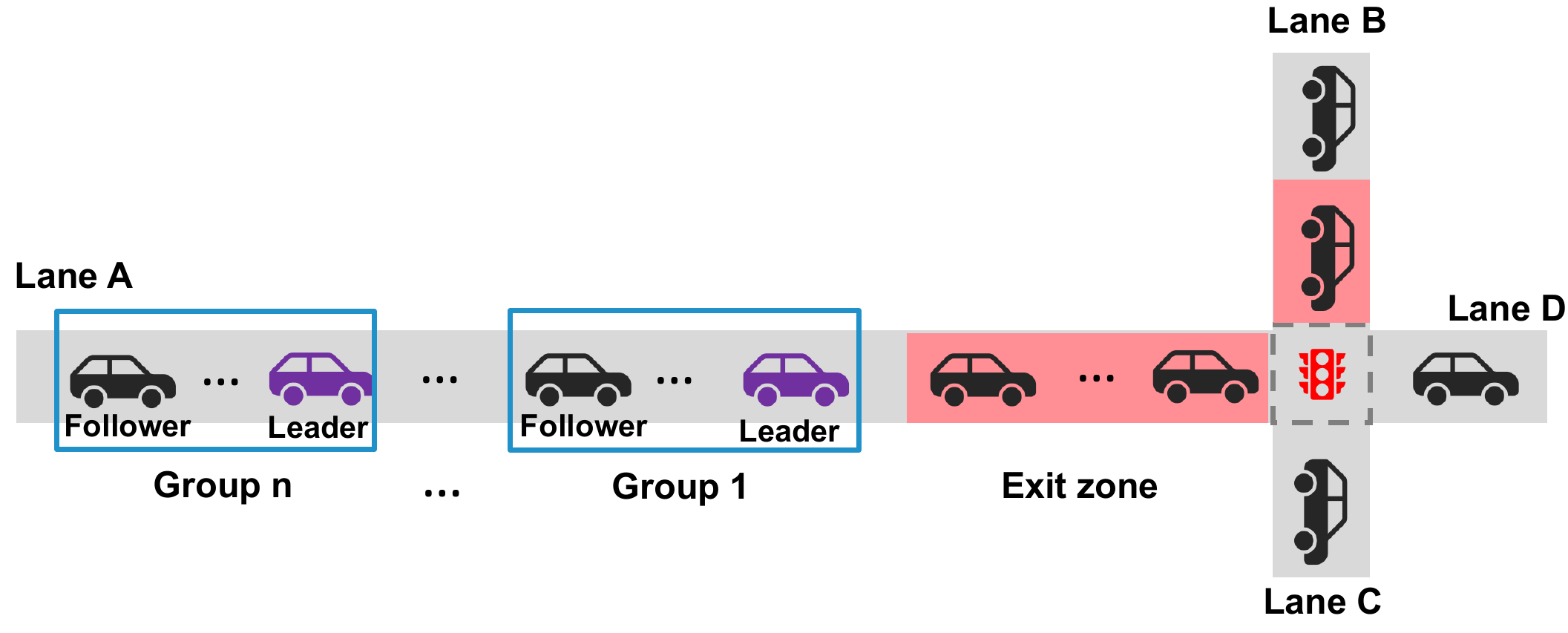}
 \caption{Group design}
 \label{fig:group_design}
\end{figure}

\subsection{Vehicle Grouping}
\label{subsec:grouping}
\sys performs its grouping judiciously, only considering grouping vehicles when they appear to be consistently moving slowly, relative to the speed limit of their lanes. 

Such a criterion requires the examination of a sequence of vehicles. 
While a lane seems a natural container where we can assess the average vehicle speed, lanes (especially long ones) often have uneven congestion situations, especially with traffic lights. 

\sys takes an ad-hoc approach that takes into consideration both vehicle speed and lane length. 
For each lane, we first define an \textit{exit zone} at the end of the lane where vehicles, if grouped, will ``escape'' and return to the original, individual SUMO simulation.  
The length of this exit zone is set relative to the length of the corresponding lane (currently configured at 10\% of the latter), while bounded by a configurable cap (by default 50 meters in \sys). 
This allows vehicles, under congestion, to disband in time to handle the upcoming junction, including changing lanes if necessary. 
Note that we do not enforce a lower limit on exit zone length. 
This is based on the observation that very short edges (road segments) are unlikely to have multiple lanes in the first place, and the group leader always follow the full SUMO individual vehicle simulation process. 

Figure~\ref{fig:group_design} illustrates a lane under heavy congestion, with the exit zone marked facing the destination junction. 
For the remaining lane region, we partition them into $k$ zones, with vehicles within each zone examined collectively.
In our current implementation, we empirically set $k$ at 3. 
The rationale behind is that when there is heavy congestion, we hope to achieve large groups on long lanes, while on short lanes we need more agile reaction to traffic changes and lane ends. 

Within each lane zone, at the beginning of each timestep, \sys computes the average speed of the vehicles. 
It identify the zone as \textit{congested}, if this average speed is under $\alpha\!\cdot\!S$, where $S$ is the speed limit of the current edge.
$\alpha$ is a configurable parameter.
Our evaluation sets this threshold conservatively, at 0, meaning that we only turn on \sys's grouping when a zone have cars not moving at all in the previous timestep.  

\subsection{Vehicle Simulation under Grouping}
\label{subsec:copying}
If a lane zone is marked as congested, the vehicles currently within this zone forms a virtual group. 
Within each group, the foremost vehicle naturally becomes the \textit{leader}, who undergoes normal SUMO simulation.
The other members of the group, called \textit{followers}, each has its \texttt{leader} attribute set appropriately.
For leaders or vehicles not under grouping, this attribute remains NULL. 

At each simulation timestep, SUMO goes through several major steps in updating a vehicle's state:
(1) \emph{plan move} (where it computes safe speeds for all vehicles for the next a few lanes, and registers approaching vehicle information for all incoming lanes (which are expected to emerge on the current lane), 
(2) \emph{set junction approaches} (where it registers junction approaches based on planned speeds as basis for right-of-way decision), (3) \emph{execute movement} (where it decides right-of-way and executes movements), and finally
(4) \emph{change lanes} (where it processes potential lane changes). 
Our profiling shows that over 90\% of the simulation time is spent on these four steps. 
\sys adds a preceding step, where it examines the grouping status of the vehicle being updated. 

Based on the check result, leaders and ungrouped (individual) vehicles move on with business as usual. 
The followers, on the other hand, skip most of the first three steps listed above (except for limited metadata updates), as they do not need sophisticated environment assessment and reaction under congestion. 
Instead, they directly move to step (4), where they simply copy the updated speed from their leaders, and adjust their locations accordingly. 

The formed groups are re-examined per timestep. 
A group is disbanded under either of two circumstances: it fails the congestion speed requirement, or its leader steps into the exit zone. 
Rather than shifting vehicles across groups, \sys simply recalculates the grouping for the whole lane, which we found to contribute to less than 1\% of the total per-timestep simulation overhead.

\section{Evaluation}
\label{sec:evaluation}

\subsection{Experiment Setup}
\label{subsec:setup}
\para{Testbed}
We conduct all of our experiments on an AWS EC2 c6g.xlarge instance running on Ubuntu 18.04.  
It has 32 ARM cores, 64GB DRAM, and 32MB L3 cache.
Our baseline uses the latest SUMO version (master branch as of June 2020).\footnote{https://github.com/eclipse/sumo}

\begin{figure}[htb]
    \centering
    \subfloat[Corniche on satellite map]{\includegraphics[width=0.23\textwidth]{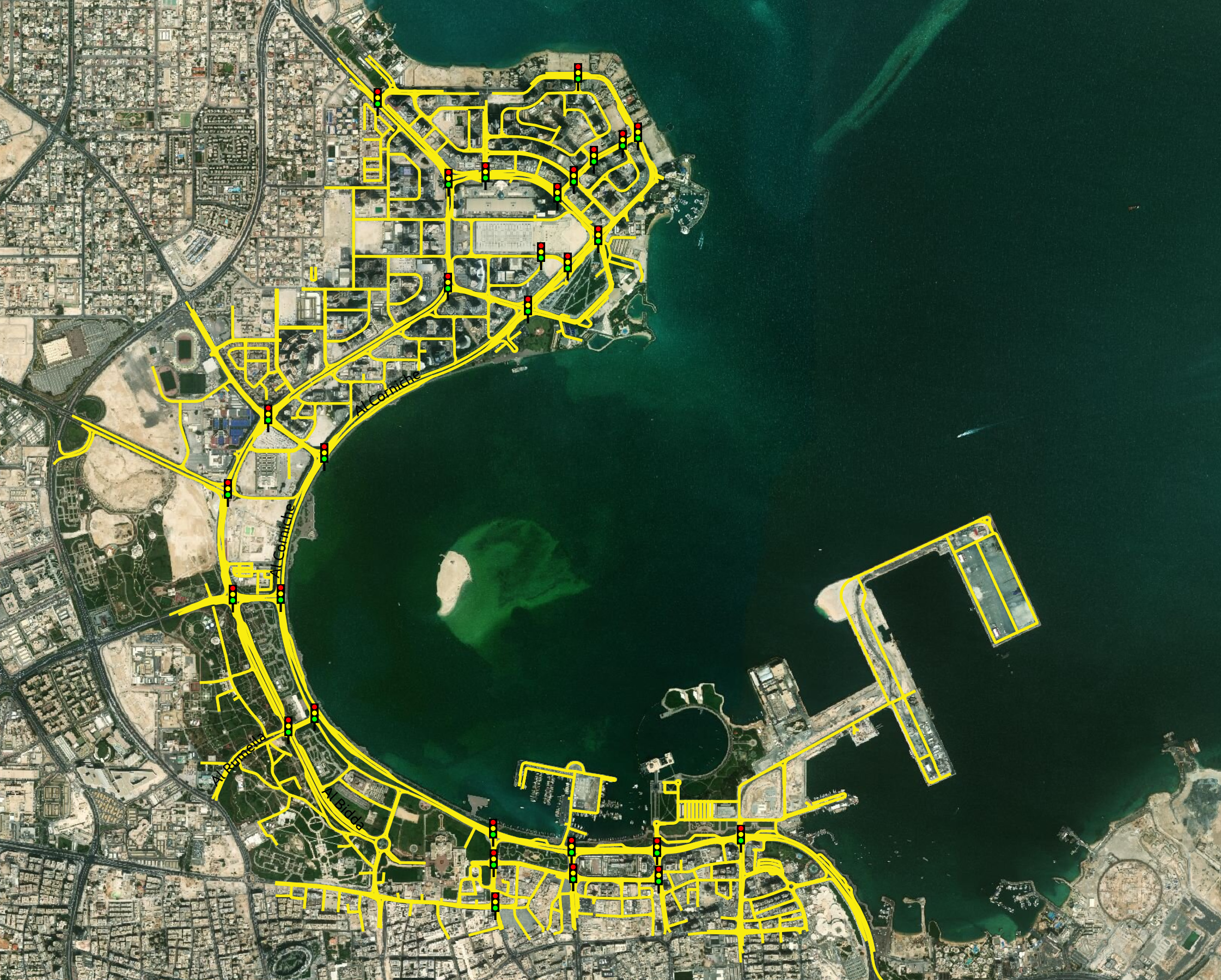}\label{fig:corniche_map}}%
    \qquad
    \subfloat[Corniche in SUMO]{\includegraphics[width=0.20\textwidth]{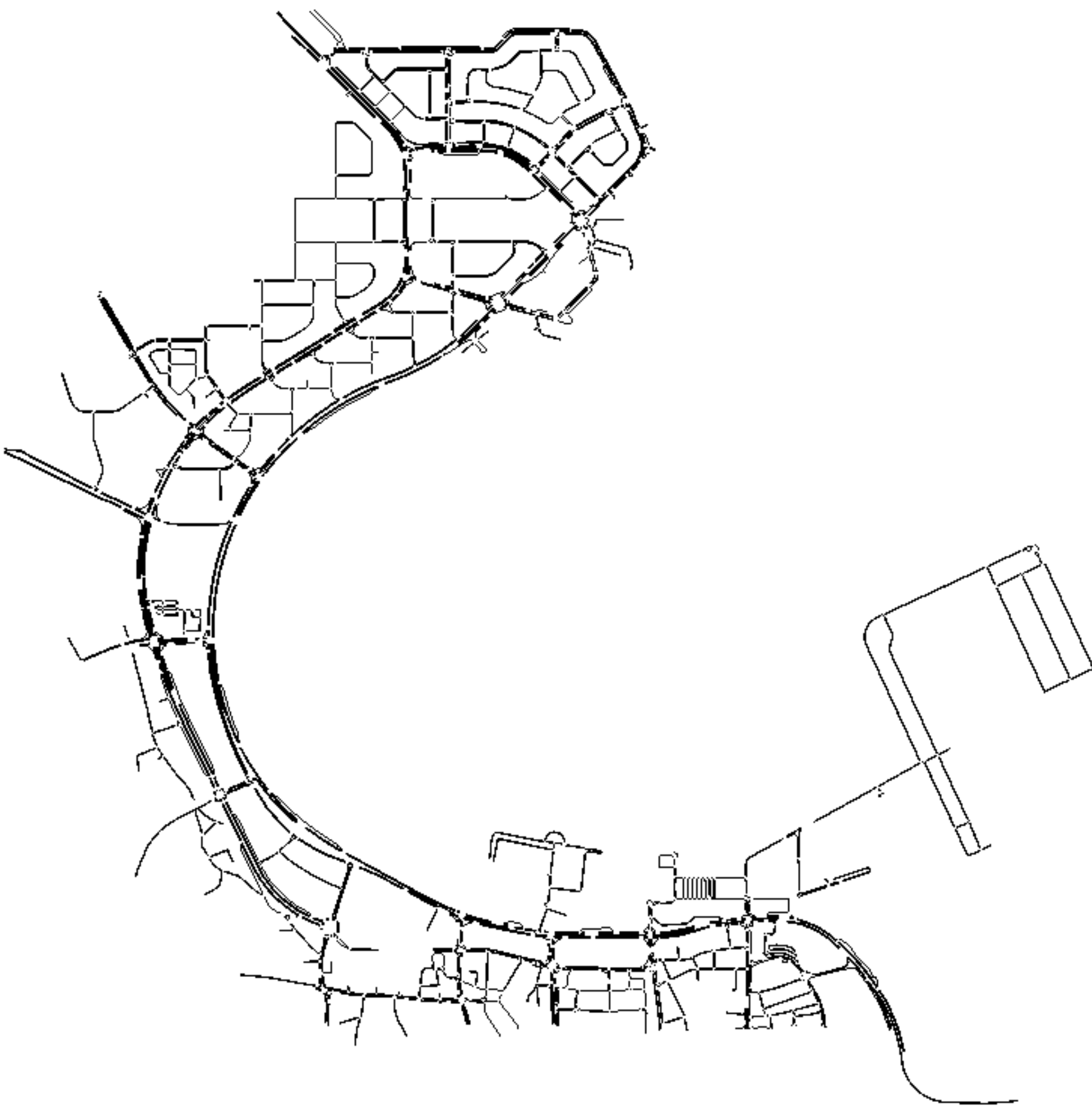}\label{fig:corniche_sumo}}%
    
    \subfloat[Cologne network]{\includegraphics[width=0.35\textwidth]{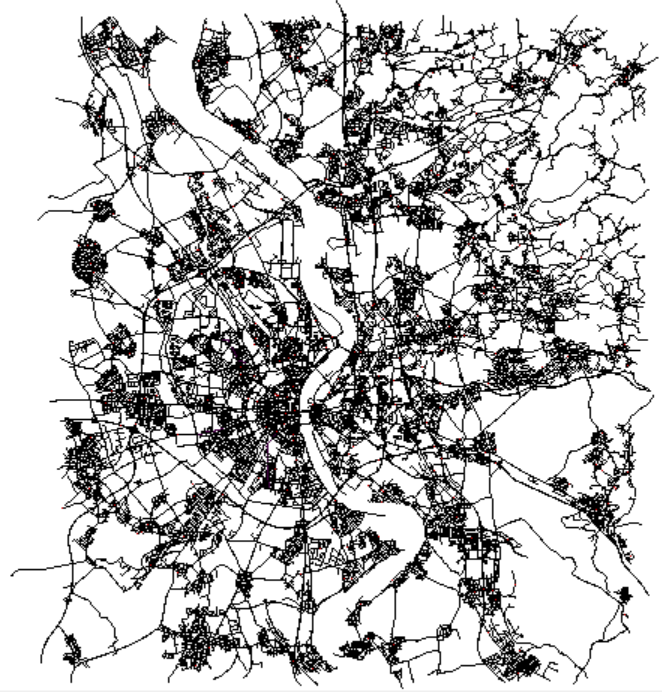}\label{fig:cologne}}
        
    \subfloat[Synthetic grid network]{\includegraphics[width=0.45\textwidth]{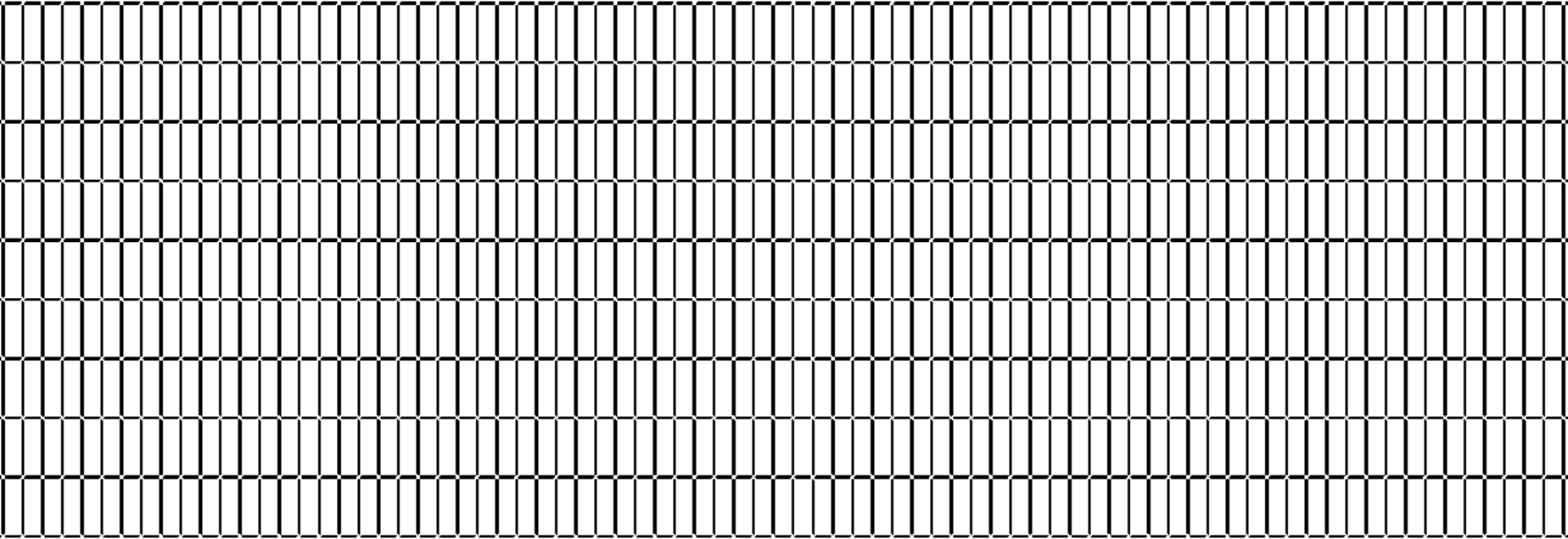}\label{fig:grid}}
    \label{fig:corniche}%
    
    \caption{Networks used: Doha Corniche, Cologne, and synthetic grid}%
    \label{fig:corniche}%
\end{figure}

\para{Road networks and traffic} 
We use both real and synthetic road networks in our development testing and performance evaluation. 

Figure~\ref{fig:corniche_map} and \ref{fig:corniche_sumo} show the aforementioned Corniche network (satellite map on the left and SUMO-visualized network on the right). 
The area surrounds the 7.8-km long Al Corniche street (3 lanes in each direction) in Doha and the network we used contains its busiest section, within Doha's West Bay (``downtown'' area). 
This section contains access to key institutes (such as the Qatar Parliament, the Ministry of Interior, the Ministry of Public Health, the National Mosque), major tourist attractions (such as the famed Museum of Islamic Art, the Souq market, and the scenic Corniche Promenade), and many hotels.
Right outside of the network section we have other visitor hot-spots: the Pearl and the Katara Cultural Village at the north and the new Qatar National Museum at the south. 
Its prominent location and frequent congestion conditions have made the Corniche area an important target in transportation management for the upcoming Qatar 2022 FIFA World Cup event. 
The event also poses unprecedented traffic challenges, as it will be the first time in World Cup history where all matches are hosted by a single city. 

This SUMO road network (Figure~\ref{fig:corniche_sumo}) includes the Corniche road and smaller streets joining it, containing 1186 junctions, 5661 edges (7931 lanes in total), and 29 traffic signals. 
The lanes (counting only a single direction of each road segment) add up to 193 kilometers. 

\cb{Our parallel simulation scalability test also used a much larger Cologne city network (Figure~\ref{fig:cologne}) available from the SUMO website,\footnote{https://sumo.dlr.de/docs/Data/Scenarios/TAPASCologne.html} with 82399 lanes and 11454 km total length, with synthetic traffic.}

To adjust network parameters and scale, we also follow the common practice of using grid networks. 
Figure~\ref{fig:grid} illustrates a sample grid map used in our evaluation, which roughly approximates the size and shape of Manhattan. It contains a 150x10 grid, with edge length of 100m (horizontal) and 300m (vertical) respectively. 

We generate synthetic traffic on the two road networks with SUMO's built-in tools. 
First we use \texttt{randomTrips.py} to generate random Original-Destination (OD) pairs,
then \texttt{duarouter} to generate routes from the OD pairs. By varying the car insertion rate, we control the traffic intensity and congestion levels.

In all the experiments we use a fine-grain time step of 0.5 seconds.



\subsection{QarSUMO Parallelization}


\begin{figure}[htb]
 \centering
 \includegraphics[width=0.3\textwidth]{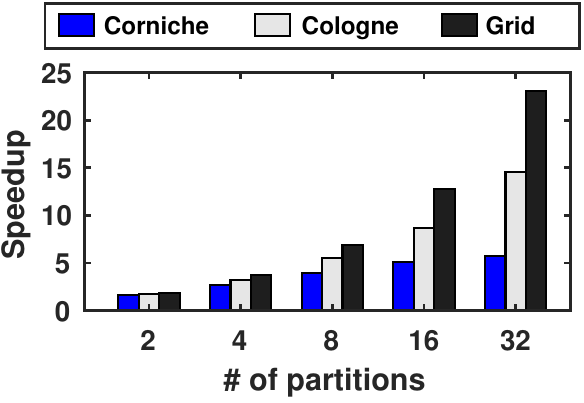}
 \caption{Performance scalability of \sys parallel execution. \sys is able to significantly reduce simulation time with parallelization, while the speedup achieved depends on the road network and traffic.}
\label{fig:parallel_speedup}
\end{figure}

We start by assessing \sys's parallel simulation performance. 
Figure~\ref{fig:parallel_speedup} shows the speedup over native SUMO's sequential execution. 
Here we give strong scalability results, where the road network and traffic is fixed while we increase the number of network partitions (and \sys processes). 

\begin{figure}[htb]
 \centering
 \includegraphics[width=0.48\textwidth]{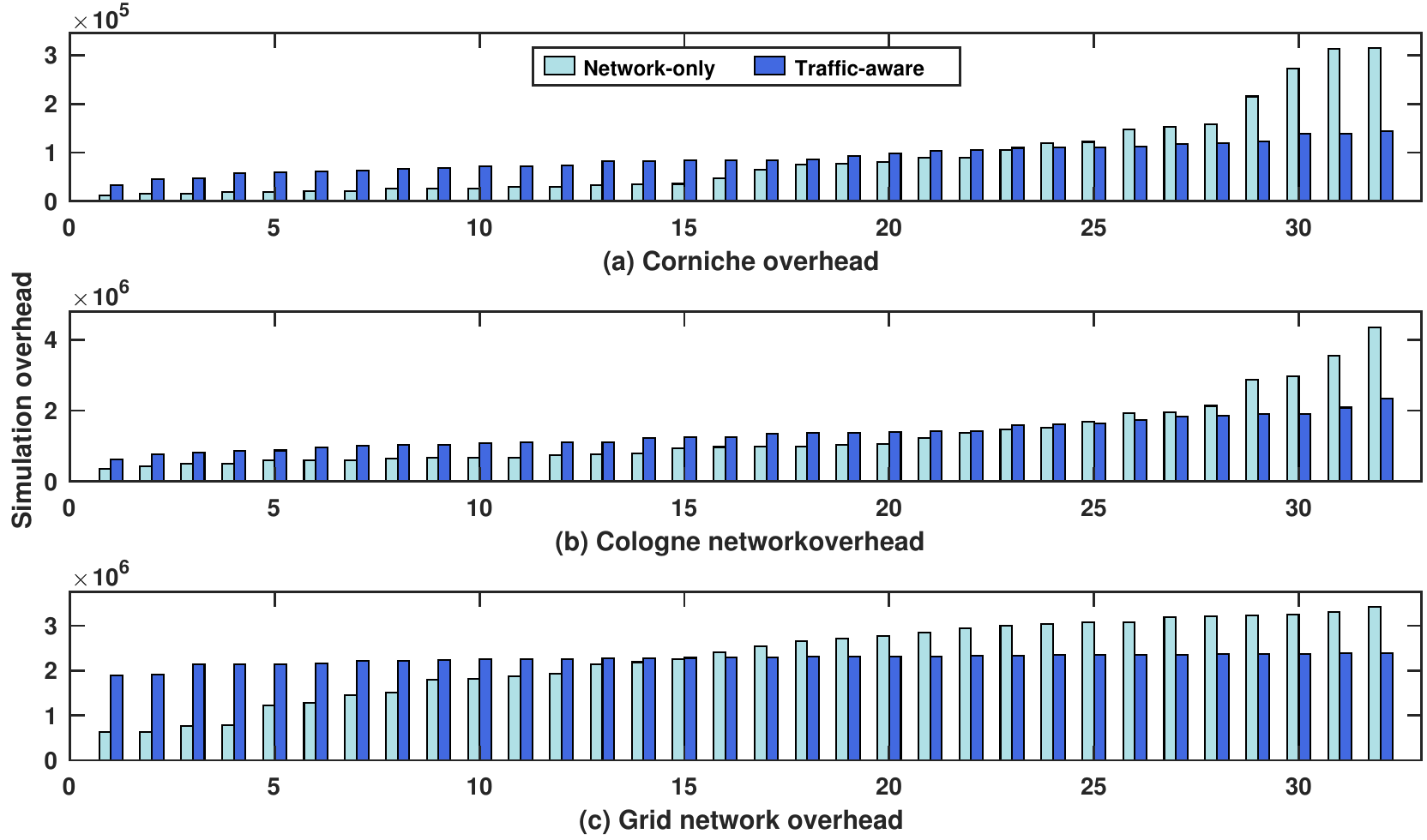}
 \caption{\cb{Traffic-aware partitioning results in significant reduction in simulation
 overhead, for all three networks at 32 partitions.}}
 \label{fig:workload_distribution}
\end{figure}

With the larger and more regular grid network, \sys maintains its efficiency relatively well, achieving a speedup of 23.05 at 32 partitions.
With the much less regular Corniche network, however, \sys's performance improves little beyond 8 partitions, reaching only 5.70 at 32.
The reason is both due to the complex road layout and the relatively small size of the Corniche graph. 
Both contribute to making load balancing harder, while the latter also makes synchronization overhead occupying higher weight in overall computation. 
\cb{With the also irregular, but significantly larger Cologne city network, \sys's scalability is between the above two cases, achieving a speedup of 14.6 at 32 partitions.
}

\cb{We confirm our analysis by examining the load balance situation with 32 partitions, for all three networks. }
Figure~\ref{fig:workload_distribution} shows the (sorted) per-partition total simulation overhead, calculated as the total number of simulation steps each partition performed (one simulation step per vehicle per timestep). 
Compared with the static partitioning, where METIS is fed with only road network topology for communication cost minimization, the traffic-aware partitioning adopted by \sys does significantly improve load balance across the partitions for both networks. 
\cb{Meanwhile, it is evident when comparing the Corniche, Cologne,
and grid network 
results, the former two have much heavier load imbalance, even after traffic-aware partitioning.
This reveals challenges in simulation load balancing with irregular real-world networks, which is a topic for future study. }

\begin{table}[htb]
\begin{tabular}{|c|c|c|c|c|c|}
\hline
Vehicles   & 3600 & 7200 & 14400 & 28800 \\ \hline
Message Size (MB) & 27.41 & 89.14 & 377.17 & 828.30  \\ 
Communication Time (\%) & 6.91\% & 3.24\% & 1.24\% & 0.71\% \\ \hline
\end{tabular}
\vspace{5pt}
\caption{\cb{Communication overhead, measured by message size and the percentage of time spent on message passing, evaluated on Corniche network with different traffic scale}}
\label{table:communication_overhead}
\end{table}

\cb{Next we assess the communication overhead. 
Table~\ref{table:communication_overhead} gives the total message size as well as the fraction of simulation time spent on inter-partition message passing, for a 32-partition simulation on Corniche \cb{(our least scalable network)} with variable number of vehicles.}
Intuitively, the total message size scales up as more vehicles are simulated. 
\cb{Note that the increase appears beyond linear: when the number of vehicles reaches 28800 ($8\times$), the transmitted message size increases by $30\times$. 
This is due to that the increased vehicle volumes slow down traffic, requiring more simulation steps for the vehicles to finish their trips and consequently producing higher total message volumes.
}

\cb{The relative cost of communication, however, remains low and actually decreases with the vehicle volume.
With 3600 vehicles, \sys spends only $6.91\%$ of its total execution time on synchronization and communication via MPI, with 27.41MB messages transmitted in total.
With 28800 vehicles, such relative overhead of message passing drops to $0.71\%$, thanks to our aggressive message batching: more messages get aggregated to the same destination partition, making communication cheaper.}

\begin{figure}[htb]
 \centering
 \includegraphics[width=0.32\textwidth]{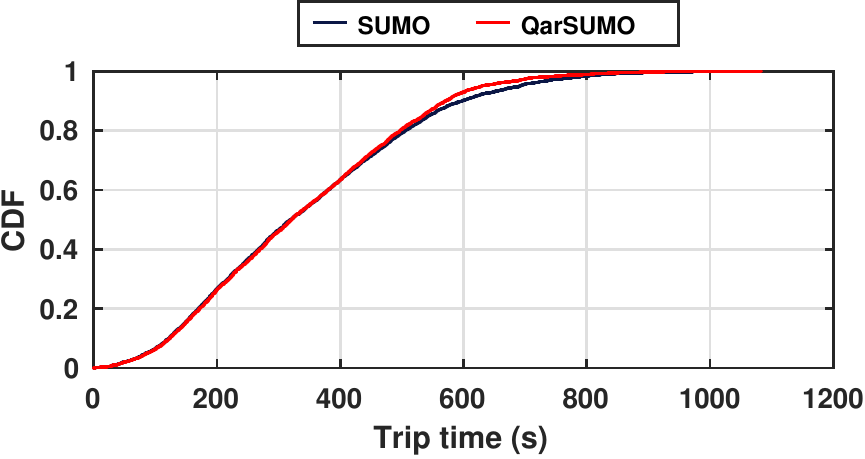}
 \caption{Trip time distribution comparison on Corniche network. The distributions are very
 similar except a slight increase for \sys for higher trip times. } 
 \label{fig:eval_parallel_accuracy}
\end{figure}

Figure~\ref{fig:eval_parallel_accuracy} depicts sample accuracy results of a 32-partition Corniche traffic simulation. 
Here we give the CDF of all the vehicles' trip time (time to travel from origin to destination), according to the simulation results by SUMO and \sys. 
As shown in the figure, the systems generate highly similar trip time distributions. 
The minor differences in simulation results are due to the imperfection of the synchronization across the border edges.
For example, the state of the primary vehicle of one timestep can only be propagated to the shadow vehicle in the next step, resulting in one timestep latency (0.5 seconds in our case), which could get accumulated as a vehicle traverses more and more border edges. 

\begin{table}[htb]
\begin{tabular}{|c|c|c|c|c|c|}
\hline
 Network/Partition   & 2 & 4  & 8  & 16 & 32 \\ \hline
Corniche    & 2.07\% & 2.45\% & 4.65\% & 4.87\% & 5.46\% \\ 
Cologne    &  1.82\% & 1.97\% & 2.10\% & 2.85\% & 2.53\% \\ 
Grid   & 0.91\% &  1.03\% &  1.26\%  & 1.45\% & 1.94\% \\ \hline
\end{tabular}
\vspace{5pt}
\caption{\cb{Relative trip time difference between SUMO and \sys}}
\label{table:parallel_accuracy}
\end{table}

Table~\ref{table:parallel_accuracy} gives more detailed accuracy evaluation results. 
For each test, we compared the trip time of individual vehicles simulated by SUMO and \sys, and calculate the relative difference as their absolute difference divided by the SUMO time. 
For the Corniche network, the errors starts small, and grow larger as the number of partitions increase, but mostly remain under 5\%.
The aforementioned tiny discrepancies due to the 0.5-second insertion delay would be accumulated, and render larger differences once in a while at a traffic light. 
The grid network, on the other hand, is a larger network, rendering lower ratio of edges being border edges, and reduces the overall relative impact of the ``simulation drift''. 
\cb{As expected, the Cologne network also has accuracy results between Corniche and grid.}
It is among our immediate future work items to investigate solutions to further reduce parallel simulation inaccuracies. 

\subsection{QarSUMO Simulation Under Congestion}
\label{subsec:congst-results}
Next, we evaluate the effectiveness of our simulation optimization under congestion. 

\begin{figure}[htb]
 \centering
 \includegraphics[width=0.48\textwidth]{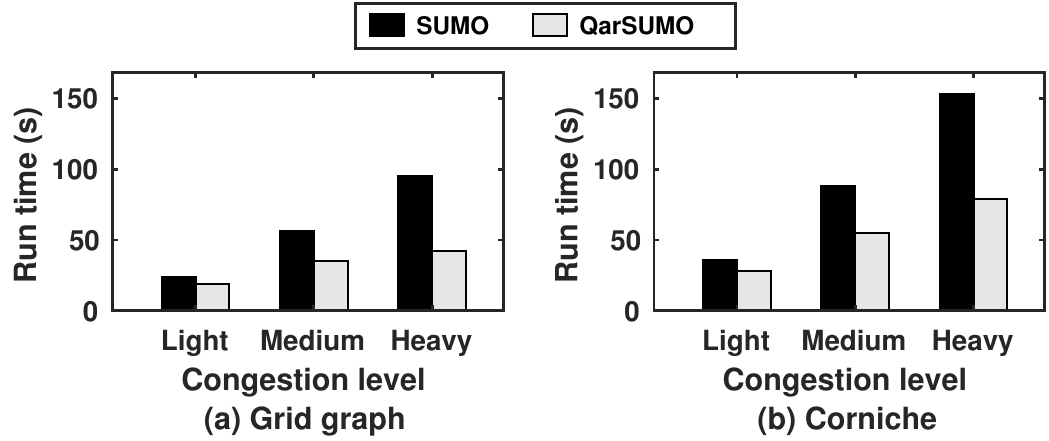}
 \caption{Virtual grouping results in a substantial saving in simulation time for both
 the Grid and Corniche networks.}
 \label{fig:eval_grouping}
\end{figure}
Figure~\ref{fig:eval_grouping} gives the execution time of both networks using synthetic traffic that generated different levels of congestion. 
As it is hard to generate the exact matching levels of congestion across the two networks, we produce three levels of traffic for each network. 
For Corniche, the ``light'', ``medium'', and ``heavy'' traffic correspond to 40.37\%, 64.82\%, and 80.33\% of vehicle-steps satisfying \sys's strict grouping criterion (0 speed and not within exit zones), respectively. 
For the grid network, the three levels of traffic have 42.95\%, 63.16\%, and 82.59\% of vehicle-steps meeting the threshold. 

Compared with the native SUMO, with its virtual grouping optimization, \sys significantly saves the simulation time by reducing detailed, per-vehicle updates under congestion.
As expected, the saving grows as congestion intensifies, obtaining a speedup of 1.93$\times$ with Corniche, and 2.26$\times$ with the grid network. 

\begin{figure}[htb]
 \centering
 \includegraphics[width=0.48\textwidth]{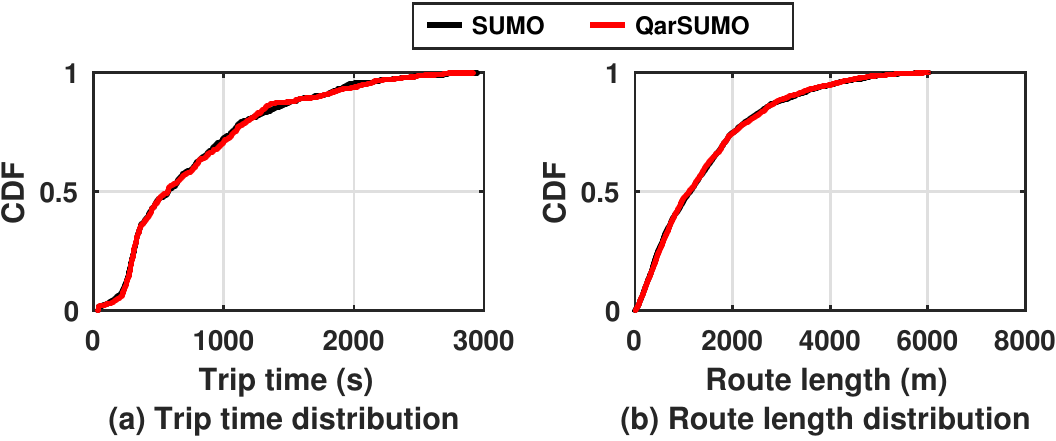}
 \caption{Trip time (for arrived vehicles) and travel distance (for en-route vehicles) CDFs}
 \label{fig:eval_congest_accuracy}
\end{figure}
Figure~\ref{fig:eval_congest_accuracy} gives a high-level accuracy comparison between SUMO and \sys. 
As we add congestion to the experiments and evaluate the congested segments of the simulation runs, many vehicles have not arrived at their destination at the end of the simulation. 
For the vehicles that have arrived, we draw the CDF of their trip time (Figure~\ref{fig:eval_congest_accuracy}(a));
for those stuck on the way, we draw the CDF of their traveled distance 
(Figure~\ref{fig:eval_congest_accuracy}(b)).
Again \sys generates nearly identical distributions as by the native SUMO. 

\begin{table}[htb]
\resizebox{\columnwidth}{!}{
\begin{tabular}{|c|c|c|c|c|}
\hline
                          & Congestion level        & Slight  & Medium  & Heavy   \\ \hline
\multirow{2}{*}{Grid} & Trip time difference    & 4.39\% & 6.22\% & 6.22\% \\ \cline{2-5} 
                          & Route length difference & 2.28\% & 8.80\% & 8.80\% \\ \hline
\multirow{2}{*}{Corniche}     & Trip time difference    & 2.98\% & 2.55\% & 2.55\% \\ \cline{2-5} 
                          & Travel distance difference & 1.43\% & 1.18\% & 1.18\% \\ \hline                          
\end{tabular}
}
\vspace{5pt}
\caption{Congestion optimization accuracy}
\label{table:congestion_accuracy}
\end{table}

Table~\ref{table:congestion_accuracy} calculates the per-vehicle relative difference between SUMO and \sys, for the above two types of vehicles, each sorted by trip time or travel distance. 
In the majority of cases, the relative difference is within 3\%. 
Like in the parallel execution evaluation, we see higher errors with the larger and signal-dense grid network, where it is easier for slight differences to accumulate into larger per-vehicle diversions. 

\subsection{Overall Performance}
\label{subsec:eval_overall}
Finally, we evaluate the combined \sys, with both parallel execution and grouping optimization. 
Here we perform typical training sample data generation for reinforcement learning (RL), simulating a one-hour episode on both networks. 
We use a moderate congestion level (between 30\% and 40\% of vehicles meeting our congestion/grouping criteria), which are of great interest to RL-based policy optimization. 
Such training typically require hundreds to thousands of such simulation-generated episodes. 

\begin{figure}[htb]
 \centering
 \includegraphics[width=0.48\textwidth]{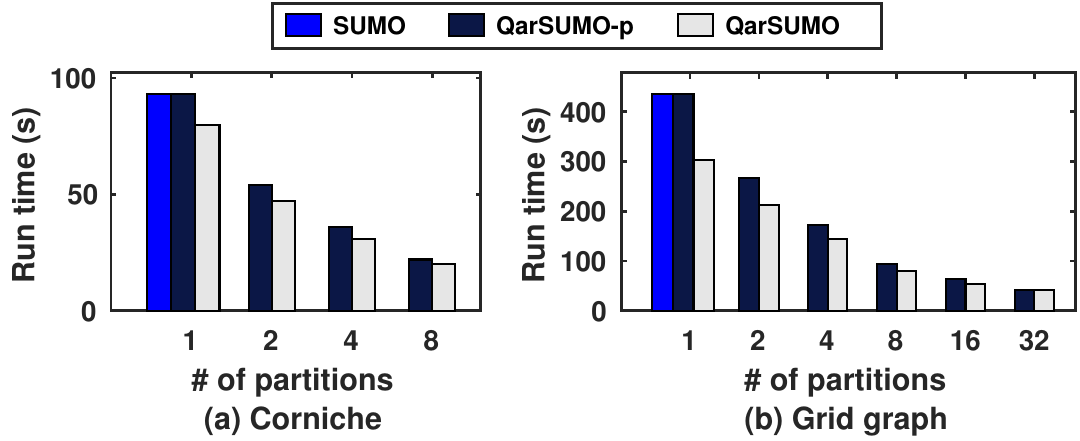}
 \caption{Overall \sys performance in episode generation for RL}
 \label{fig:eval_rl}
\end{figure}

Figure\ref{fig:eval_rl} gives execution time results, where \sys-p denotes \sys with parallelization but not grouping. 
Considering results shown \sys leveling off beyond 8 partitions with Corniche, here we only test 2 to 8 partitions. 
We see that \sys delivers considerable speedup (up to 4.22$\times$ with Corniche and 10.60$\times$ with grid) by parallelization alone. 

On top of that, \sys's congestion simulation optimization brings additional time saving (15-30\% simulation time reduction from \sys-p). 
First, here we only have medium congestion level, with under 40\% of simulation steps meeting the speed threshold for grouping. 
Second, with parallelization, the overall simulation time is dominated by the slowest partition for each timestep, making only congestion saving to this particular partition visible.

Together, the two techniques reduces the overall one-hour episode generation time by 4.65$\times$ for Corniche, and 10.61$\times$ for grid, greatly saving the cost of RL training. 
For example, the time to generate 500 hours of continuous episodes for the grid network  will be reduced from 60.41 hours to 5.70 hours. 

\begin{table}[htb]
\begin{tabular}{|c|c|c|c|c|c|}
\hline
 Network/Partition   & 2 & 4  & 8  & 16 & 32 \\ \hline
Corniche    & 2.07\% & 8.21\% & 6.6\% & -  & -  \\ 
Grid   & 0.93\% &  1.02\% &  1.11\%  & 1.37\% & 2.03\% \\ \hline
\end{tabular}
\vspace{5pt}
\caption{Relative route length difference between SUMO and \sys, in one-hour episode generation}
\label{table:rl_accuracy}
\end{table}
Table~\ref{table:rl_accuracy} summarizes accuracy evaluation results, using the same method as in Table~\ref{table:parallel_accuracy}. 
Again the grid network delivers higher accuracy (error below 2.03\%), while the smaller Corniche network has larger errors (up to 8.21\%) with 4 or 8 partitions. 
In our future work, we plan to evaluate the impact of such approximated simulation on RL learning efficiency.

\section{Related Work} 
\label{sec:related-work}


Traffic simulation is an important and active topic within transportation research. Prominent traffic
simulators that are used both in practice and research include SUMO~\cite{lopez2018microscopic}, VISSIM~\cite{fellendorf2010microscopic}, ParamGrid~\cite{klefstad2005distributed}, MATSim~\cite{w2016multi}, SMARTS~\cite{ramamohanarao2016smarts}
and GeoSparkSim~\cite{fu2019demonstrating}. The SMARTS project~\cite{ramamohanarao2016smarts} proposed a taxonomy of features to compare different 
simulators including support for distributed computing and decentralized synchronization. 
The official release of SUMO does not support distributed computing and thus cannot support large-scale simulations.

Two prominent systems that do enable distributed traffic simulation are SMARTS~\cite{ramamohanarao2016smarts} and
GeoSparkSim~\cite{fu2019demonstrating}. 
SMARTS supports master-slave computation model as well as decentralized synchronization, where workers only communicate with other workers that share a road edge. 
In order to ensure that each simulation timestep finishes
within similar time within different slaves, SMARTS uses a workload
distribution strategy that creates a number of workers among whom the workload can be nearly evenly distributed. 
Recall that due to the Car Following Model (CFM) and the Lane Changing Model (LCM), essential to most traffic simulation software, 
workers will have to coordinate amongst each other when two consecutive cars on a road network end up being assigned to different workers.

GeoSparkSim~\cite{fu2019demonstrating} is a recent proposal
that extends GeoSpark~\cite{yu2015geospark} to support
traffic simulation. 
GeoSparkSim relaxes a key assumption in SMARTS that the spatial distributions of vehicles remain the same during the simulation. GeoSparkSim introduces VehicleRDD, an in-memory extension of
RDDs (Resilient Distributed Dataset) from Spark~\cite{zaharia2010spark},
which also records spatio-temporal information of vehicles. 
As the vehicular traffic moves in the network the VehicleRDDs get transformed to maintain a spatio-temporal workload balance. 

Several attempts have been made to address the lack of multi-threaded and distributed computing support in SUMO, such as Spartsim~\cite{ventresque2012spartsim} and dSUMO~\cite{bragard2013dsumo}. 
Like \sys, the core idea of these methods is also based on network partitioning. 
However in both proposals the overhead of keeping the compute nodes synchronized and maintaining a balanced workload outstrips the inherent advantages of distributed computing.
\sys also uses METIS for network partitioning, but provides better scalability than reported by the above systems as it reduces the inter-partition synchronization overhead.
It also avoids expensive dynamic load balancing by performing traffic-aware network partitioning using the available traffic profile.

\cb{
There is also a parallel implementation of the TRANSIMS micro-simulation tool~\cite{nagel2001parallel}.
It uses cellular automata (CA) to model driving dynamics (while SUMO adopts the more powerful continuous model). 
Like \sys, it also adopts METIS-based network partitioning and edge-based inter-partition communication. 
It is able to improve load balancing by collecting load statistics from a prior run, targeting the scenario that similar traffic situations are repeatedly simulated. 
\sys, instead, can take into account routing information for the current simulation for its load balancing. 
}

Another recent study~\cite{arroyo2018new} parallelizing SUMO only targets grid networks and cuts the network horizontally or vertically to smaller rectangles.
Similarly, the recently developed Cityflow traffic simulator, also targeting reinforcement learning training, is also limited to grid networks.
\sys, on the other hand, tackles real world networks which often have irregular shapes, which makes the synchronization much more challenging.

Besides parallelization, higher microscopic simulation efficiency has also been achieved by considering the varying congestion level in different areas and switching from a constant time-step, queue-based simulation to an event-driven simulation~\cite{charypar2006event}. 
Approaches like this aimed at reducing the computation time in very low traffic areas where few traffic events occur. 
In contrast, QarSUMO proposes a novel technique to optimize the computation time in highly congested areas, which account for the heaviest workload in SUMO. 
To our knowledge, our virtual grouping method is the first in reducing simulation cost for congested traffic. 

\section{Conclusion}
\label{sec:conclusion}
In this paper, we present \sys, an enhancement to the  widely used SUMO open-source traffic simulation software. 
\sys enables SUMO to take advantage of powerful multi-core server nodes, by adding high-level parallelization outside of its simulation kernel. 
It also improves SUMO's simulation efficiency under severe traffic congestions, by reducing vehicle update granularity. 
Our evaluation with real and synthetic networks confirm that \sys is able to bring a speedup of up to a factor of five 
over the native SUMO performance in realistic scenarios, while maintaining reasonable simulation accuracy. 

Our immediate future work plan includes comprehensive testing and code cleaning, as well as further simulation accuracy improvement, to prepare for \sys's open-source release.
Also, we are working on pre-processing the full traffic network for Doha, to enable much larger scale (in space and time), city-wide traffic simulation using \sys. 
For future research, we plan to investigate dynamic load balancing (without network re-partitioning) and more detailed/flexible congestion detection schemes, as well as further simulation optimization designed specifically for reinforcement learning. 
\section{Acknowledgments}
\cb{We thank the program chairs and our reviewers, for their invaluable help and insightful comments.
We appreciate the patient clarification and helpful suggestions from the SUMO developer team during this project.
We would also like to thank Italconsult S.p.A, who on behalf of the Qatar Ministry of Transport and Communications (MOTC), provided us with the Doha Corniche network data and support.
The co-first-authors Hao Chen and Ke Yang are both supported by the QCRI research internship during this work. 
Phillip Taylor's 2019 summer internship at QCRI was through Princeton University’s International Internship Program and we thank Eman Fituri for facilitating our hosting project. 
}

\bibliographystyle{ACM-Reference-Format}
\bibliography{main}


\begin{thebibliography}{29}


\ifx \showCODEN    \undefined \def \showCODEN     #1{\unskip}     \fi
\ifx \showDOI      \undefined \def \showDOI       #1{#1}\fi
\ifx \showISBNx    \undefined \def \showISBNx     #1{\unskip}     \fi
\ifx \showISBNxiii \undefined \def \showISBNxiii  #1{\unskip}     \fi
\ifx \showISSN     \undefined \def \showISSN      #1{\unskip}     \fi
\ifx \showLCCN     \undefined \def \showLCCN      #1{\unskip}     \fi
\ifx \shownote     \undefined \def \shownote      #1{#1}          \fi
\ifx \showarticletitle \undefined \def \showarticletitle #1{#1}   \fi
\ifx \showURL      \undefined \def \showURL       {\relax}        \fi
\providecommand\bibfield[2]{#2}
\providecommand\bibinfo[2]{#2}
\providecommand\natexlab[1]{#1}
\providecommand\showeprint[2][]{arXiv:#2}

\bibitem[\protect\citeauthoryear{??}{sum}{2020a}]%
        {sumo:projects}
 \bibinfo{year}{2020}\natexlab{a}.
\newblock \bibinfo{booktitle}{\emph{SUMO Projects Page}}.
\newblock
\urldef\tempurl%
\url{https://sumo.dlr.de/docs/Other/Projects.html}
\showURL{%
\tempurl}


\bibitem[\protect\citeauthoryear{??}{sum}{2020b}]%
        {sumoconfwebsite}
 \bibinfo{year}{2020}\natexlab{b}.
\newblock \bibinfo{booktitle}{\emph{SUMO User Conference}}.
\newblock
\urldef\tempurl%
\url{https://www.eclipse.org/sumo/conference/}
\showURL{%
\tempurl}


\bibitem[\protect\citeauthoryear{Arroyo, Acosta, Espinosa, and Espinosa}{Arroyo
  et~al\mbox{.}}{2018}]%
        {arroyo2018new}
\bibfield{author}{\bibinfo{person}{Nicol{\'a}s Arroyo},
  \bibinfo{person}{Andr{\'e}s Acosta}, \bibinfo{person}{Jairo Espinosa}, {and}
  \bibinfo{person}{Jorge Espinosa}.} \bibinfo{year}{2018}\natexlab{}.
\newblock \showarticletitle{A new strategy for synchronizing traffic flow on a
  distributed simulation using SUMO}.
\newblock \bibinfo{journal}{\emph{EPiC Series in Engineering}}
  \bibinfo{volume}{2} (\bibinfo{year}{2018}), \bibinfo{pages}{152--161}.
\newblock


\bibitem[\protect\citeauthoryear{{Bragard}, {Ventresque}, and Murphy}{{Bragard}
  et~al\mbox{.}}{2013}]%
        {bragard2013dsumo}
\bibfield{author}{\bibinfo{person}{Quentin {Bragard}}, \bibinfo{person}{Anthony
  {Ventresque}}, {and} \bibinfo{person}{{B.E.} Murphy, Liam}.}
  \bibinfo{year}{2013}\natexlab{}.
\newblock \showarticletitle{dSUMO: towards a distributed SUMO}.
\newblock \bibinfo{journal}{\emph{The first SUMO User Conference, Berlin,
  Germany, 15 - 17 May, 2013}} (\bibinfo{year}{2013}).
\newblock


\bibitem[\protect\citeauthoryear{Charypar, Axhausen, and Nagel}{Charypar
  et~al\mbox{.}}{2006}]%
        {charypar2006event}
\bibfield{author}{\bibinfo{person}{David Charypar}, \bibinfo{person}{Kay~W
  Axhausen}, {and} \bibinfo{person}{Kai Nagel}.}
  \bibinfo{year}{2006}\natexlab{}.
\newblock \showarticletitle{An event-driven queue-based microsimulation of
  traffic flow}.
\newblock \bibinfo{journal}{\emph{Arbeitsberichte Verkehrs-und Raumplanung}}
  \bibinfo{volume}{406} (\bibinfo{year}{2006}).
\newblock


\bibitem[\protect\citeauthoryear{Fellendorf and Vortisch}{Fellendorf and
  Vortisch}{2010}]%
        {fellendorf2010microscopic}
\bibfield{author}{\bibinfo{person}{Martin Fellendorf} {and}
  \bibinfo{person}{Peter Vortisch}.} \bibinfo{year}{2010}\natexlab{}.
\newblock \showarticletitle{Microscopic traffic flow simulator VISSIM}.
\newblock In \bibinfo{booktitle}{\emph{Fundamentals of traffic simulation}}.
  \bibinfo{publisher}{Springer}, \bibinfo{pages}{63--93}.
\newblock


\bibitem[\protect\citeauthoryear{Fu, Yu, and Sarwat}{Fu et~al\mbox{.}}{2019a}]%
        {fu2019building}
\bibfield{author}{\bibinfo{person}{Zishan Fu}, \bibinfo{person}{Jia Yu}, {and}
  \bibinfo{person}{Mohamed Sarwat}.} \bibinfo{year}{2019}\natexlab{a}.
\newblock \showarticletitle{Building a large-scale microscopic road network
  traffic simulator in apache spark}. In \bibinfo{booktitle}{\emph{2019 20th
  IEEE International Conference on Mobile Data Management (MDM)}}. IEEE,
  \bibinfo{pages}{320--328}.
\newblock


\bibitem[\protect\citeauthoryear{Fu, Yu, and Sarwat}{Fu et~al\mbox{.}}{2019b}]%
        {fu2019demonstrating}
\bibfield{author}{\bibinfo{person}{Zishan Fu}, \bibinfo{person}{Jia Yu}, {and}
  \bibinfo{person}{Mohamed Sarwat}.} \bibinfo{year}{2019}\natexlab{b}.
\newblock \showarticletitle{Demonstrating geosparksim: A scalable microscopic
  road network traffic simulator based on Apache spark}. In
  \bibinfo{booktitle}{\emph{Proceedings of the 16th International Symposium on
  Spatial and Temporal Databases}}. \bibinfo{pages}{186--189}.
\newblock


\bibitem[\protect\citeauthoryear{Gabriel, Fagg, Bosilca, Angskun, Dongarra,
  Squyres, Sahay, Kambadur, Barrett, Lumsdaine, et~al\mbox{.}}{Gabriel
  et~al\mbox{.}}{2004}]%
        {gabriel2004open}
\bibfield{author}{\bibinfo{person}{Edgar Gabriel}, \bibinfo{person}{Graham~E
  Fagg}, \bibinfo{person}{George Bosilca}, \bibinfo{person}{Thara Angskun},
  \bibinfo{person}{Jack~J Dongarra}, \bibinfo{person}{Jeffrey~M Squyres},
  \bibinfo{person}{Vishal Sahay}, \bibinfo{person}{Prabhanjan Kambadur},
  \bibinfo{person}{Brian Barrett}, \bibinfo{person}{Andrew Lumsdaine},
  {et~al\mbox{.}}} \bibinfo{year}{2004}\natexlab{}.
\newblock \showarticletitle{Open MPI: Goals, concept, and design of a next
  generation MPI implementation}. In \bibinfo{booktitle}{\emph{European
  Parallel Virtual Machine/Message Passing Interface Users’ Group Meeting}}.
  Springer, \bibinfo{pages}{97--104}.
\newblock


\bibitem[\protect\citeauthoryear{Gazis, Herman, and Rothery}{Gazis
  et~al\mbox{.}}{1961}]%
        {gazis}
\bibfield{author}{\bibinfo{person}{Denos~C. Gazis}, \bibinfo{person}{Robert
  Herman}, {and} \bibinfo{person}{Richard~W. Rothery}.}
  \bibinfo{year}{1961}\natexlab{}.
\newblock \showarticletitle{Nonlinear Follow-The-Leader Models of Traffic
  Flow}.
\newblock \bibinfo{journal}{\emph{Operations Research}} \bibinfo{volume}{9},
  \bibinfo{number}{4} (\bibinfo{year}{1961}), \bibinfo{pages}{545--567}.
\newblock
\showISSN{0030364X, 15265463}


\bibitem[\protect\citeauthoryear{Karypis and Kumar}{Karypis and Kumar}{1998}]%
        {karypis1998fast}
\bibfield{author}{\bibinfo{person}{George Karypis} {and} \bibinfo{person}{Vipin
  Kumar}.} \bibinfo{year}{1998}\natexlab{}.
\newblock \showarticletitle{A fast and high quality multilevel scheme for
  partitioning irregular graphs}.
\newblock \bibinfo{journal}{\emph{SIAM Journal on scientific Computing}}
  \bibinfo{volume}{20}, \bibinfo{number}{1} (\bibinfo{year}{1998}),
  \bibinfo{pages}{359--392}.
\newblock


\bibitem[\protect\citeauthoryear{Kheterpal, Parvate, Wu, Kreidieh, Vinitsky,
  and Bayen}{Kheterpal et~al\mbox{.}}{2018}]%
        {kheterpal2018flow}
\bibfield{author}{\bibinfo{person}{Nishant Kheterpal}, \bibinfo{person}{Kanaad
  Parvate}, \bibinfo{person}{Cathy Wu}, \bibinfo{person}{Aboudy Kreidieh},
  \bibinfo{person}{Eugene Vinitsky}, {and} \bibinfo{person}{Alexandre Bayen}.}
  \bibinfo{year}{2018}\natexlab{}.
\newblock \showarticletitle{Flow: Deep reinforcement learning for control in
  sumo}.
\newblock \bibinfo{journal}{\emph{EPiC Series in Engineering}}
  \bibinfo{volume}{2} (\bibinfo{year}{2018}), \bibinfo{pages}{134--151}.
\newblock


\bibitem[\protect\citeauthoryear{Klefstad, Zhang, Lai, Jayakrishnan, and
  Lavanya}{Klefstad et~al\mbox{.}}{2005}]%
        {klefstad2005distributed}
\bibfield{author}{\bibinfo{person}{Raymond Klefstad}, \bibinfo{person}{Yue
  Zhang}, \bibinfo{person}{Mingjie Lai}, \bibinfo{person}{R Jayakrishnan},
  {and} \bibinfo{person}{Riju Lavanya}.} \bibinfo{year}{2005}\natexlab{}.
\newblock \showarticletitle{A distributed, scalable, and synchronized framework
  for large-scale microscopic traffic simulation}. In
  \bibinfo{booktitle}{\emph{Proceedings. 2005 IEEE Intelligent Transportation
  Systems, 2005.}} IEEE, \bibinfo{pages}{813--818}.
\newblock


\bibitem[\protect\citeauthoryear{Krajzewicz, Hertkorn, R{\"o}ssel, and
  Wagner}{Krajzewicz et~al\mbox{.}}{2002}]%
        {krajzewicz2002sumo}
\bibfield{author}{\bibinfo{person}{Daniel Krajzewicz}, \bibinfo{person}{Georg
  Hertkorn}, \bibinfo{person}{Christian R{\"o}ssel}, {and}
  \bibinfo{person}{Peter Wagner}.} \bibinfo{year}{2002}\natexlab{}.
\newblock \showarticletitle{SUMO (Simulation of Urban MObility)-an open-source
  traffic simulation}. In \bibinfo{booktitle}{\emph{Proceedings of the 4th
  middle East Symposium on Simulation and Modelling (MESM20002)}}.
  \bibinfo{pages}{183--187}.
\newblock


\bibitem[\protect\citeauthoryear{Kuyer, Whiteson, Bakker, and Vlassis}{Kuyer
  et~al\mbox{.}}{2008}]%
        {2016_multiagent_reinforcement_learning_coordination_graphs}
\bibfield{author}{\bibinfo{person}{Lior Kuyer}, \bibinfo{person}{Shimon
  Whiteson}, \bibinfo{person}{Bram Bakker}, {and} \bibinfo{person}{Nikos
  Vlassis}.} \bibinfo{year}{2008}\natexlab{}.
\newblock \showarticletitle{Multiagent Reinforcement Learning for Urban Traffic
  Control Using Coordination Graphs}. In \bibinfo{booktitle}{\emph{Machine
  Learning and Knowledge Discovery in Databases}},
  \bibfield{editor}{\bibinfo{person}{Walter Daelemans}, \bibinfo{person}{Bart
  Goethals}, {and} \bibinfo{person}{Katharina Morik}} (Eds.).
  \bibinfo{publisher}{Springer Berlin Heidelberg}, \bibinfo{address}{Berlin,
  Heidelberg}, \bibinfo{pages}{656--671}.
\newblock
\showISBNx{978-3-540-87479-9}


\bibitem[\protect\citeauthoryear{Lopez, Behrisch, Bieker-Walz, Erdmann,
  Fl{\"o}tter{\"o}d, Hilbrich, L{\"u}cken, Rummel, Wagner, and WieBner}{Lopez
  et~al\mbox{.}}{2018}]%
        {lopez2018microscopic}
\bibfield{author}{\bibinfo{person}{Pablo~Alvarez Lopez},
  \bibinfo{person}{Michael Behrisch}, \bibinfo{person}{Laura Bieker-Walz},
  \bibinfo{person}{Jakob Erdmann}, \bibinfo{person}{Yun-Pang
  Fl{\"o}tter{\"o}d}, \bibinfo{person}{Robert Hilbrich},
  \bibinfo{person}{Leonhard L{\"u}cken}, \bibinfo{person}{Johannes Rummel},
  \bibinfo{person}{Peter Wagner}, {and} \bibinfo{person}{Evamarie WieBner}.}
  \bibinfo{year}{2018}\natexlab{}.
\newblock \showarticletitle{Microscopic traffic simulation using sumo}. In
  \bibinfo{booktitle}{\emph{2018 21st International Conference on Intelligent
  Transportation Systems (ITSC)}}. IEEE, \bibinfo{pages}{2575--2582}.
\newblock


\bibitem[\protect\citeauthoryear{Mousavi, Schukat, and Howley}{Mousavi
  et~al\mbox{.}}{2017}]%
        {2017_PG_and_QLearning_intersection_image}
\bibfield{author}{\bibinfo{person}{Seyed~Sajad Mousavi},
  \bibinfo{person}{Michael Schukat}, {and} \bibinfo{person}{Enda Howley}.}
  \bibinfo{year}{2017}\natexlab{}.
\newblock \showarticletitle{Traffic light control using deep policy-gradient
  and value-function-based reinforcement learning}.
\newblock \bibinfo{journal}{\emph{IET Intelligent Transport Systems}}
  \bibinfo{volume}{11}, \bibinfo{number}{7} (\bibinfo{year}{2017}),
  \bibinfo{pages}{417--423}.
\newblock


\bibitem[\protect\citeauthoryear{Nagel and Rickert}{Nagel and Rickert}{2001}]%
        {nagel2001parallel}
\bibfield{author}{\bibinfo{person}{Kai Nagel} {and} \bibinfo{person}{Marcus
  Rickert}.} \bibinfo{year}{2001}\natexlab{}.
\newblock \showarticletitle{Parallel implementation of the TRANSIMS
  micro-simulation}.
\newblock \bibinfo{journal}{\emph{Parallel Comput.}} \bibinfo{volume}{27},
  \bibinfo{number}{12} (\bibinfo{year}{2001}), \bibinfo{pages}{1611--1639}.
\newblock


\bibitem[\protect\citeauthoryear{Ramamohanarao, Xie, Kulik, Karunasekera,
  Tanin, Zhang, and Khunayn}{Ramamohanarao et~al\mbox{.}}{2016}]%
        {ramamohanarao2016smarts}
\bibfield{author}{\bibinfo{person}{Kotagiri Ramamohanarao},
  \bibinfo{person}{Hairuo Xie}, \bibinfo{person}{Lars Kulik},
  \bibinfo{person}{Shanika Karunasekera}, \bibinfo{person}{Egemen Tanin},
  \bibinfo{person}{Rui Zhang}, {and} \bibinfo{person}{Eman~Bin Khunayn}.}
  \bibinfo{year}{2016}\natexlab{}.
\newblock \showarticletitle{Smarts: Scalable microscopic adaptive road traffic
  simulator}.
\newblock \bibinfo{journal}{\emph{ACM Transactions on Intelligent Systems and
  Technology (TIST)}} \bibinfo{volume}{8}, \bibinfo{number}{2}
  (\bibinfo{year}{2016}), \bibinfo{pages}{1--22}.
\newblock


\bibitem[\protect\citeauthoryear{Rizzo, Vantini, and Chawla}{Rizzo
  et~al\mbox{.}}{2019}]%
        {rizzo2019time}
\bibfield{author}{\bibinfo{person}{Stefano~Giovanni Rizzo},
  \bibinfo{person}{Giovanna Vantini}, {and} \bibinfo{person}{Sanjay Chawla}.}
  \bibinfo{year}{2019}\natexlab{}.
\newblock \showarticletitle{Time Critic Policy Gradient Methods for Traffic
  Signal Control in Complex and Congested Scenarios}. In
  \bibinfo{booktitle}{\emph{Proceedings of the 25th ACM SIGKDD International
  Conference on Knowledge Discovery \& Data Mining}}.
  \bibinfo{pages}{1654--1664}.
\newblock


\bibitem[\protect\citeauthoryear{Sokolowski and Banks}{Sokolowski and
  Banks}{2011}]%
        {sokolowski2011principles}
\bibfield{author}{\bibinfo{person}{John~A Sokolowski} {and}
  \bibinfo{person}{Catherine~M Banks}.} \bibinfo{year}{2011}\natexlab{}.
\newblock \bibinfo{booktitle}{\emph{Principles of modeling and simulation: a
  multidisciplinary approach}}.
\newblock \bibinfo{publisher}{John Wiley \& Sons}.
\newblock


\bibitem[\protect\citeauthoryear{Tang, Naphade, Liu, Yang, Birchfield, Wang,
  Kumar, Anastasiu, and Hwang}{Tang et~al\mbox{.}}{2019}]%
        {tang2019cityflow}
\bibfield{author}{\bibinfo{person}{Zheng Tang}, \bibinfo{person}{Milind
  Naphade}, \bibinfo{person}{Ming-Yu Liu}, \bibinfo{person}{Xiaodong Yang},
  \bibinfo{person}{Stan Birchfield}, \bibinfo{person}{Shuo Wang},
  \bibinfo{person}{Ratnesh Kumar}, \bibinfo{person}{David Anastasiu}, {and}
  \bibinfo{person}{Jenq-Neng Hwang}.} \bibinfo{year}{2019}\natexlab{}.
\newblock \showarticletitle{Cityflow: A city-scale benchmark for multi-target
  multi-camera vehicle tracking and re-identification}. In
  \bibinfo{booktitle}{\emph{Proceedings of the IEEE Conference on Computer
  Vision and Pattern Recognition}}. \bibinfo{pages}{8797--8806}.
\newblock


\bibitem[\protect\citeauthoryear{Ventresque, Bragard, Liu, Nowak, Murphy,
  Theodoropoulos, and Liu}{Ventresque et~al\mbox{.}}{2012}]%
        {ventresque2012spartsim}
\bibfield{author}{\bibinfo{person}{Anthony Ventresque},
  \bibinfo{person}{Quentin Bragard}, \bibinfo{person}{Elvis~S Liu},
  \bibinfo{person}{Dawid Nowak}, \bibinfo{person}{Liam Murphy},
  \bibinfo{person}{Georgios Theodoropoulos}, {and} \bibinfo{person}{Qi Liu}.}
  \bibinfo{year}{2012}\natexlab{}.
\newblock \showarticletitle{Spartsim: A space partitioning guided by road
  network for distributed traffic simulations}. In
  \bibinfo{booktitle}{\emph{2012 IEEE/ACM 16th International Symposium on
  Distributed Simulation and Real Time Applications}}. IEEE,
  \bibinfo{pages}{202--209}.
\newblock


\bibitem[\protect\citeauthoryear{W~Axhausen, Horni, and Nagel}{W~Axhausen
  et~al\mbox{.}}{2016}]%
        {w2016multi}
\bibfield{author}{\bibinfo{person}{Kay W~Axhausen}, \bibinfo{person}{Andreas
  Horni}, {and} \bibinfo{person}{Kai Nagel}.} \bibinfo{year}{2016}\natexlab{}.
\newblock \bibinfo{booktitle}{\emph{The multi-agent transport simulation
  MATSim}}.
\newblock \bibinfo{publisher}{Ubiquity Press}.
\newblock


\bibitem[\protect\citeauthoryear{Wei, Chen, Zheng, Wu, Gayah, Xu, and Li}{Wei
  et~al\mbox{.}}{2019a}]%
        {wei2019presslight}
\bibfield{author}{\bibinfo{person}{Hua Wei}, \bibinfo{person}{Chacha Chen},
  \bibinfo{person}{Guanjie Zheng}, \bibinfo{person}{Kan Wu},
  \bibinfo{person}{Vikash Gayah}, \bibinfo{person}{Kai Xu}, {and}
  \bibinfo{person}{Zhenhui Li}.} \bibinfo{year}{2019}\natexlab{a}.
\newblock \showarticletitle{PressLight: Learning Max Pressure Control to
  Coordinate Traffic Signals in Arterial Network}. In
  \bibinfo{booktitle}{\emph{Proceedings of the 25th ACM SIGKDD International
  Conference on Knowledge Discovery \& Data Mining}}.
  \bibinfo{pages}{1290--1298}.
\newblock


\bibitem[\protect\citeauthoryear{Wei, Zheng, Gayah, and Li}{Wei
  et~al\mbox{.}}{2019b}]%
        {wei2019survey}
\bibfield{author}{\bibinfo{person}{Hua Wei}, \bibinfo{person}{Guanjie Zheng},
  \bibinfo{person}{Vikash Gayah}, {and} \bibinfo{person}{Zhenhui Li}.}
  \bibinfo{year}{2019}\natexlab{b}.
\newblock \bibinfo{title}{A Survey on Traffic Signal Control Methods}.
\newblock
\newblock
\showeprint[arxiv]{1904.08117}~[cs.LG]


\bibitem[\protect\citeauthoryear{Wei, Zheng, Yao, and Li}{Wei
  et~al\mbox{.}}{2018}]%
        {intellilight}
\bibfield{author}{\bibinfo{person}{Hua Wei}, \bibinfo{person}{Guanjie Zheng},
  \bibinfo{person}{Huaxiu Yao}, {and} \bibinfo{person}{Zhenhui Li}.}
  \bibinfo{year}{2018}\natexlab{}.
\newblock \showarticletitle{IntelliLight: A Reinforcement Learning Approach for
  Intelligent Traffic Light Control}. In \bibinfo{booktitle}{\emph{Proceedings
  of the 24th ACM SIGKDD International Conference on Knowledge Discovery
  \&\#38; Data Mining}} (London, United Kingdom) \emph{(\bibinfo{series}{KDD
  '18})}. \bibinfo{publisher}{ACM}, \bibinfo{address}{New York, NY, USA},
  \bibinfo{pages}{2496--2505}.
\newblock
\showISBNx{978-1-4503-5552-0}


\bibitem[\protect\citeauthoryear{Yu, Wu, and Sarwat}{Yu et~al\mbox{.}}{2015}]%
        {yu2015geospark}
\bibfield{author}{\bibinfo{person}{Jia Yu}, \bibinfo{person}{Jinxuan Wu}, {and}
  \bibinfo{person}{Mohamed Sarwat}.} \bibinfo{year}{2015}\natexlab{}.
\newblock \showarticletitle{Geospark: A cluster computing framework for
  processing large-scale spatial data}. In
  \bibinfo{booktitle}{\emph{Proceedings of the 23rd SIGSPATIAL International
  Conference on Advances in Geographic Information Systems}}.
  \bibinfo{pages}{1--4}.
\newblock


\bibitem[\protect\citeauthoryear{{Zaharia}, {Chowdhury}, {Franklin}, {Shenker},
  and {Stoica}}{{Zaharia} et~al\mbox{.}}{2010}]%
        {zaharia2010spark}
\bibfield{author}{\bibinfo{person}{Matei {Zaharia}}, \bibinfo{person}{Mosharaf
  {Chowdhury}}, \bibinfo{person}{Michael~J. {Franklin}}, \bibinfo{person}{Scott
  {Shenker}}, {and} \bibinfo{person}{Ion {Stoica}}.}
  \bibinfo{year}{2010}\natexlab{}.
\newblock \showarticletitle{Spark: cluster computing with working sets}. In
  \bibinfo{booktitle}{\emph{HotCloud'10 Proceedings of the 2nd USENIX
  conference on Hot topics in cloud computing}}. \bibinfo{pages}{10--10}.
\newblock


\end{thebibliography}

\end{document}